\documentclass[lettersize,journal]{IEEEtran}
\usepackage{amsmath,amsfonts}
\usepackage{algorithmic}
\usepackage{algorithm}
\usepackage{array}
\usepackage{textcomp}
\usepackage{stfloats}
\usepackage{url}
\usepackage{verbatim}
\usepackage{graphicx}
\usepackage{cite}

\usepackage{amsfonts}
\usepackage{amsmath}
\usepackage{color}
\usepackage{subfigure}
\usepackage{bm}
\newtheorem{theorem}{Theorem}

\begin{document}

\title{Wasserstein Dependent Graph Attention Network for Collaborative Filtering with Uncertainty}

\author{Haoxuan Li, Yuanxin Ouyang, Zhuang Liu, Wenge Rong, Zhang Xiong
        % <-this % stops a space

\thanks{Manuscript accepted on June 24, 2024. This work supported by the National Natural Science Foundation of China (No.62377002).}% <-this % stops a space

\thanks{Haoxuan Li, Yuanxin Ouyang, Zhuang Liu, Wenge Rong and Zhang Xiong are with Engineering Research Center of Advanced Computer Application Technology, Ministry of Education, and School of Computer Science and Engineering, Beihang University, 100191, Beijing, China. (e-mail: sy2206129@buaa.edu.cn; oyyx@buaa.edu.cn; liuzhuang@buaa.edu.cn; w.rong@buaa.edu.cn;
xiongz@buaa.edu.cn).
}

}

\markboth{Journal of \LaTeX\ Class Files,~Vol.~14, No.~8, August~2021}%
{Shell \MakeLowercase{\textit{et al.}}: A Sample Article Using IEEEtran.cls for IEEE Journals}


\maketitle

\begin{abstract}
Collaborative filtering (CF) is an essential technique in recommender systems that provides personalized recommendations by only leveraging user-item interactions. However, most CF methods represent users and items as fixed points in the latent space, lacking the ability to capture uncertainty. While probabilistic embedding is proposed to intergrate uncertainty, they suffer from several limitations when introduced to graph-based recommender systems.
Graph convolutional network framework would confuse the semantic of uncertainty in the nodes, and  similarity measured by Kullback–Leibler (KL) divergence suffers from degradation problem and demands an exponential number of samples.
To address these challenges, we propose a novel approach, called the \underline{W}asserstein dependent \underline{G}raph \underline{AT}tention network (W-GAT), for collaborative filtering with uncertainty. We utilize graph attention network and Wasserstein distance to learn Gaussian embedding for each user and item. Additionally, our method incorporates Wasserstein-dependent mutual information further to increase the similarity between positive pairs. Experimental results on three benchmark datasets show the superiority of W-GAT compared to several representative baselines. Extensive experimental analysis validates the effectiveness of W-GAT in capturing uncertainty by modeling the range of user preferences and categories associated with items.
\end{abstract}


\section{Introduction}
Collaborative Filtering (CF) aims to predict a user's future preference based on historical interactions, which is the most basic recommendation technology. 
Traditional CF algorithms commonly employ matrix factorization techniques to learn user and item representations, relying on inner products to compute scores between users and candidate items. Consequently, Graph Neural Networks (GNNs) have demonstrated improved capability in learning expressive representations, treating interactions between users and items as a bipartite graph. However, most methods like \cite{koren2009matrix, he2017neural, he2020lightgcn} embed users and items into low-dimensional vectors, specifically fixed points. Although they have achieved remarkable recommendation performance, they fail to capture uncertain information. Although probabilistic models like Probabilistic MF \cite{mnih2007probabilistic} and BPR \cite{rendle2012bpr} have been proposed, they only employ probability distributions as prior knowledge to facilitate the learning of more accurate deterministic embeddings. As a result, these methods fall short of explicitly addressing the challenge of modeling and incorporating uncertainty.

\begin{figure*}[htbp]
  \centering
  \includegraphics[width=\linewidth]{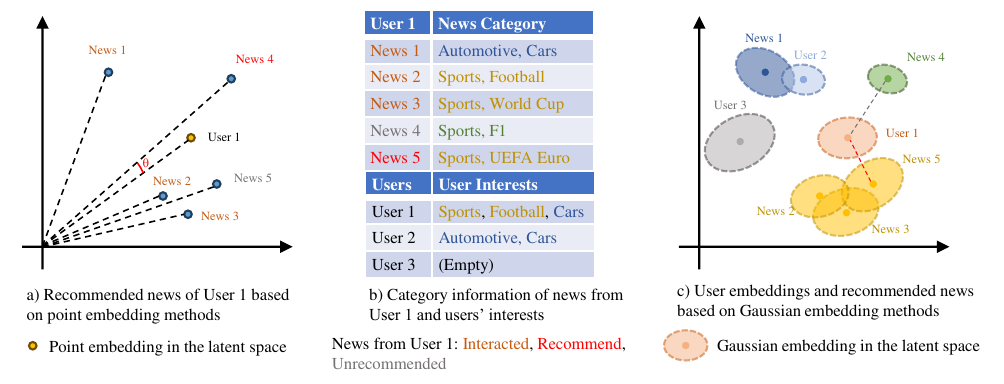}
  \caption{Differences between point embedding methods and Gaussian embedding methods.}
  \label{fig: introducation}
\end{figure*}

Representations with uncertainty have been widely employed in different fields \cite{bojchevski2017deep, fan2022sequential}. %It can effectively depict users who lack prior knowledge and capture the diversity of their interests—an advantage that fixed point representations lack. 
Illustrated in Figure \ref{fig: introducation}, point embedding methods would recommend News 4 to User 1 based on cosine similarity, failing to capture the user's genuine interest in football. While probabilistic embedding methods would recommend News 5 based on the distance of distributions. Meanwhile, it also can effectively model the uncertainty of new users (User 3) or users with multiple interests (User 1) by a more significant variance. Uncertainty in the items can also be captured by variances, where a detailed categorized item will be represented with a relatively small variance. Items with less categorical information or a large range of categories have more uncertainty and will be represented with greater variances. The uncertainty information modeled by variances gives a novel insight to generate explainable recommendations and a potential enhancement in recommendation diversity.

%For example, representing a new user with limited interaction history as a probabilistic distribution rather than a fixed point embedding could improve recommendations during the cold start period by avoiding overfitting on sparse data.

%To leverage the benefits of uncertain representation, users and items are represented as distributions in the latent space.

Moreover, recent methods aim at maximizing mutual information to improve the performances \cite{oord2018representation, wu2022effectiveness}. Mutual information portrays the amount of common information between two random variables, and intuitively, the larger the mutual information between two variables, the more information about the other variable is revealed when one random variable is given. The widely used contrastive learning loss InfoNCE \cite{oord2018representation} maximizes the mutual information between positive pairs and minimizes that between random negative pairs. And when contrastive loss is introduced, the model can capture more accurate and robust user preferences \cite{wu2022effectiveness}.

However, there are a number of challenges to tackle when using probabilistic embeddings and measuring the corresponding similarity. Firstly, when representing users and items as distributions, the message-passing GNNs would unconditionally propagate variances between users and items, despite the semantic difference in nature. Hence, more accurate ways to propagate variances across the user-item interaction graph are worth exploring.
Secondly, most approaches resort to two types of computations: Kullback–Leibler divergence (KL divergence) and Wasserstein distance to measure the similarity. However, KL divergence suffers from several disadvantages, including asymmetry, failure to satisfy the triangular inequality (i.e., transitivity), and inability to handle some cases of distribution degradation. In the CF scenario, the asymmetry of KL divergence leads to an unacceptable ordered user-item similarity\cite{fan2023mutual}. At the same time, the lack of transitivity hinders the transmission of collaborative signals through the user-item bipartite graph.

Therefore, the commonly employed KL divergence-based mutual information estimation has several limitations when dealing with probabilistic distribution embeddings. To begin with, approximating the lower bound of mutual information using KL divergence requires an exponential number of samples \cite{mcallester2020formal}. Furthermore, the mutual information computed by KL divergence inherently suffers from degradation problems, which means that the difference between two distributions is small, but the KL divergence can still be large \cite{ozair2019wasserstein}. %This results in the KL divergence between the joint probability density $p(x,y)$ and the product of the two marginal probability densities $p(x) \cdot p(y)$ becoming large, even though the difference between the two is small and intuitively, the two distributions are close to being independent, where the mutual information $I(x,y)$ should be small.
In the field of recommendation, it will inflate the mutual information between the user and irrelevant or nearly irrelevant items, introducing bias to the capture of user preferences. %\cite{courty2017learning, ozair2019wasserstein, muzellec2018generalizing} works on Wasserstein distance to tackle the limitations of KL divergence mentioned above.

To address the challenges mentioned above, we propose a novel \underline{W}asserstein dependent \underline{G}raph \underline{AT}tention network (W-GAT). Our method captures the uncertainty and adequately transmits collaborative signals by leveraging a graph attention network to learn Gaussian representations of users and items. In addition, we carefully induce Wasserstein distance to more effectively measure similarity between distributions, as well as to maximize the mutual information between users and their interacted items in order to capture more accurate user preferences. To summarize, our work makes the following contributions:
\begin{itemize}
\item We propose W-GAT, a probabilistic representation model based on the graph attention network, allowing for capturing uncertainty in users and items.

\item We introduce the Wasserstein dependency measurement into collaborative filtering to maximize mutual information between users and relevant items.

\item We conduct extensive experiments on three benchmark datasets to demonstrate the effectiveness of our proposed method over several competitive baselines.

\end{itemize}

\section{Related Work}

\textbf{Collaborative filtering} serves as a fundamental recommendation approach. In early stages, matrix factorization (MF) techniques were employed to acquire user/item representations, and inner products were used to calculate predicted scores for recommendation. As deep neural networks emerged, MF methods were integrated with neural networks to enhance the expressiveness of representations \cite{xue2017deep, he2017neural}.
Furthermore, with the advent of Graph Neural Networks (GNNs), researchers leverage GNNs to exploit the collaborative signals on user-item bipartite graphs, leading to more accurate representations and outstanding recommendation performance.
%These methods based on GNNs excel in capturing the interaction information between users and items present in the bipartite graph.
%This, in turn, results in more accurate user-item representations and outstanding recommendation performance.
NGCF \cite{wang2019neural} uses multiple graph convolutional layers on the bipartite graph to propagate embeddings to capture high-order information. Notably, LightGCN \cite{he2020lightgcn} emerged as a simplified variant of NGCF and has become the widely adopted graph encoder in recent recommender systems. %LightGCN omits the matrix feature extraction module in GCN and utilize solely on the adjacency matrix for updating node embeddings. 
Despite its simplicity, the LightGCN model serves as the backbone encoder for state-of-the-art recommendation models. Graph Attention Networks (GAT) intends to learn importance weights, where important neighbors will be given greater attention \cite{velivckovic2017graph}.

\textbf{Probability distribution embeddings} offers an effective approach to model uncertainty. Vilnis et al. \cite{vilnis2014word} represents textual information as Gaussian distributions, employing max-margin loss to capture uncertainty. KG2E \cite{he2015learning} embeds entities and relationships in knowledge graphs as Gaussians, adjusting distances between positive and negative sample pairs to enhance representation quality. 
Graph2Gauss \cite{bojchevski2017deep} calculates distribution similarity by KL divergence and optimizes the ranking loss based on energy function. 
% \cite{zhu2018deep} introduce Wasserstein distance and variational networks to propose deep variational network embedding in Wasserstein space. 
Ma et al. \cite{ma2020probabilistic} employs Wasserstein distance to measure user preferences and introduces adaptive margin loss based on user-user and item-item relationships through bilevel optimization. Fan et al. \cite{fan2022sequential} introduces the stochastic attention mechanism to the sequential recommendation, measuring similarity between historical interactions and candidate items by 2-Wasserstein distance. 
%While many methods combine additional information, such as labels and features, the exploration of collaborative filtering is still ongoing. 
%However, the distribution-based methods employed in CF are still under-explored.
Dos et al. \cite{dos2017gaussian} embeds users and items as Gaussian embeddings, utilizing the inner product between distributions as a substitute for the original vector inner product. Jiang et al. \cite{jiang2020convolutional} also adopts Gaussian distributions to characterize users and items, employing Monte-Carlo sampling and convolutional neural networks to predict interaction likelihood. 
Some Variational Autoencoder (VAE) based methods \cite{gan2023viga, ding2021semi, zhang2021wasserstein, yao2020correlated} aim to learn the underlying distributions of users and items by focusing on sampling from the learned distributions and optimizing the reconstruction loss function.

\textbf{Mutual information} has been widely employed in recent advanced recommendation methods to improve user representations and more accurately capture user preferences. The widely used contrastive learning loss, InfoNCE \cite{wu2022effectiveness}, maximizes the mutual information between positive samples to enhance recommendation effectiveness.
%Subsequent recommendation methods have built upon this foundation by incorporating techniques such as debiasing and data augmentation to further enhance recommendation outcomes \textcolor{red}{(cite relevant articles, e.g., BCloss)}.
In the field of representation learning, the Wasserstein Dependency Measure (WDM) has been proposed as an alternative to mutual information computed by KL divergence \cite{ozair2019wasserstein}. The use of Wasserstein distance addresses the limitations induced by KL divergence, which requires exponential-level sampling and will experience gradient vanishing problems during the training process. WDM achieves a more robust mutual information estimation.

\section{Methodology}
In this section, we first introduce preliminaries and problem definitions, and then we dive into several vital components of the proposed method in detail. Finally, we discuss the training loss of the model.
\subsection{Preliminaries}

Let $\mathcal{U}=\{u_1,u_2,\cdots,u_n\}$ and $\mathcal{I}=\{i_1,i_2,\cdots,i_m\}$ denote the set of users and items, respectively. We represent the user-item interactions using an undirected graph $\mathcal{G} = (\mathcal{V}, \mathcal{E})$, where $\mathcal{V}$ represents the set of vertices and is defined as $\mathcal{V} = \mathcal{U} \cup \mathcal{I}$. An edge $\{u,i\}\in\mathcal{E}$ is added to the graph if user $u$ has interacted with item $i$, where $\mathcal{E}$ represents the set of edges in the graph $\mathcal{G}$. $\mathcal{N}_u$ denotes the set of neighbors of node $u$. We aim to learn lower-dimensional Gaussian distribution embeddings of each user and item, such that users and the their interested items are similar in the hidden space given a similarity measure $\hat{y}$, $D$ is the embedding size.

\subsection{Gaussian Embeddings on Graph}
\label{sec: Gaussian on graph}

Although point embedding has yielded excellent results on recommendation tasks, it lacks the ability to mine uncertainty due to its fixed-in-space nature. To capture uncertainty, we encode each user and item as a Gaussian embedding (i.e. each element follows a Gaussian distribution), which leads to,

\begin{equation}\label{u_i_gauss_embed}
\textbf{e}_v \sim \mathcal{N}(\bm{\mu}_v,\bm{\Sigma}_v), \bm{\mu}_v \in \mathbb{R}^D, \bm{\Sigma}_v \in \mathbb{R}^{D\times D}, v\in \mathcal{V}
\end{equation}
To reduce the complexity of the model and minimize computational overhead \cite{bojchevski2017deep}, we assume that the embedding dimensions are uncorrelated. Thus, $\mathbf{\Sigma}_v$ is considered as a diagonal covariance matrix $diag(\sigma_1,\sigma_2,\cdots,\sigma_d)$ and can be further represented by a D-dimensional array.

In order to exploit the collaborative signals on the user-item interaction graph, message-passing GNN encoders are employed in the collaborative filtering model.
%LightGCN \cite{he2020lightgcn} is the most prevalent backbone graph encoder deployed on the current collaborative filtering model.
To take the most prevalent encoder LightGCN as an example, it omits the matrix feature extraction module in GCN and only utilizes the adjacency matrix for updating node embeddings, which are defined as follows,
\begin{equation} \label{lightGCN definition}
    \begin{aligned}
        & \textbf{E}^{(k+1)}=(\textbf{D}^{-\frac{1}{2}}\textbf{A}\textbf{D}^{-\frac{1}{2}})\textbf{E}^{(k)} \\
        & \textbf{E} = \frac{1}{K+1}\Sigma^K_{k=0}(\textbf{D}^{-\frac{1}{2}}\textbf{A}\textbf{D}^{-\frac{1}{2}})^k\textbf{E}^{(k)}
    \end{aligned}
\end{equation}
where $\textbf{A}$ is the adjacency matrix and $\textbf{D}$ is a diagonal matrix of node degree. $\textbf{E}$ denotes the embedding matrix and $K$ is the number of layers in LightGCN.

However, in the settings of probabilistic embedding where users and items are represented by two vectors $\bm{\mu}_i$ and $\mathbf{\Sigma}_i$, LightGCN would unconditionally propagate the node's information to its neighbors and unrestrictedly aggregate information collected from its neighbors. The variances of users capture the diversity of their interests and the uncertainty in their behavior. Conversely, items do not inherently exhibit uncertainty, but their variances represent the range of categories they belong to. Given the fundamental semantic distinction between their variances, direct propagation will result in confusion when representing users and items. We further validate this statement in Section \ref{subsubsec: limitation of LighGCN}.

Meanwhile, most recommendation methods calculate the similarity between user $u$ and candidate item $i$ by the inner product. When it comes to measuring the similarity of two multivariate Gaussian distributions, there are two widely used methods, KL divergence and 2-Wasserstein distance defined as follows,
\begin{equation}
\label{define DKL}
\begin{aligned}
    D_{KL}(\textbf{e}_u || \textbf{e}_i) = & \frac{1}{2}\Bigg( \Bigg. (\bm{\mu}_i-\bm{\mu}_u)^{T}\mathbf{\Sigma}_i^{-1}(\bm{\mu}_i-\bm{\mu}_u) + \\ & tr(\mathbf{\Sigma}_i^{-1}\mathbf{\Sigma}_u) - d + \ln{\frac{|\mathbf{\Sigma}_i|}{|\mathbf{\Sigma}_u|}}\Bigg. \Bigg)
\end{aligned}
\end{equation}
\begin{equation}
\label{define 2-W distance}
\begin{aligned}
    W_2(\textbf{e}_u, \textbf{e}_i) = ||\bm{\mu}_u-\bm{\mu}_i||^2_2 + tr\left(\mathbf{\Sigma}_u + \mathbf{\Sigma}_i - 2(\mathbf{\Sigma}_i^\frac{1}{2}\mathbf{\Sigma}_u\mathbf{\Sigma}_i^\frac{1}{2})^\frac{1}{2}\right)
\end{aligned}
\end{equation}

However, similarity measured by KL divergence suffers from several limitations. Firstly, KL divergence is asymmetric, i.e. $D_{KL}(\textbf{e}_u||\textbf{e}_i)\neq D_{KL}(\textbf{e}_i||\textbf{e}_u)$, leading to an ordered similarity between $u$ and $i$, which contradicts the symmetry of similarity in collaborative filtering. Secondly, KL divergence is sensitive to small differences and could be large even if the difference between $u$ and $i$ is small\cite{ozair2019wasserstein}. It causes the model to incorrectly exclude items that match the user's preferences from the candidate list and thereby introduce noises in capturing user preferences. Thirdly, KL divergence does not satisfy the triangle inequality, i.e., $D_{KL}(\textbf{e}_{i_1}||\textbf{e}_{i_2})\leq D_{KL}(\textbf{e}_{u}||\textbf{e}_{i_1}) + D_{KL}(\textbf{e}_{i_2}||\textbf{e}_{u})$ does not hold in general. Intuitively, if a user $u$ is similar to two items $i_1, i_2$, the distribution between $i_1$ and $i_2$ is probably similar, where KL divergence could be large on the contrary. It means KL divergence could not properly transmit the collaborative signals. Besides, when KL divergence is undefined or infinite (i.e. two distributions do not overlap), the model suffers from gradient vanishing problems, leading to an unstable training process. 

In contrast, Wasserstein distance represents the differences between two distributions regarding the actual distance between sampled data. Considering the limitations of KL divergence mentioned above, we resort to Wasserstein distance to measure similarity, which is not only symmetric but also satisfies the triangle inequality \cite{clement2008elementary}. Models also benefit from a more stable training process as Wasserstein distance provides smoother gradients when KL divergence cannot, which will be verified in Section \ref{subsubsec: limitation of MI by KL}.

\begin{figure*}[htp]
  \centering
  \includegraphics[width=\linewidth]{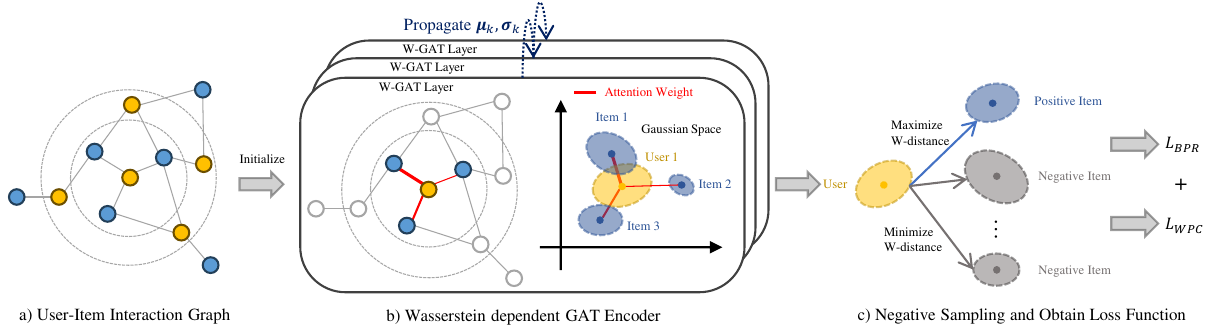}
  \caption{The framework of W-GAT. We deploy W-GAT layers which propagate the means and variances of representations on the user-item bipartite graph. Then we use the in-batch sample strategy to select positive and negative pairs. Finally, we obtain the two loss functions based on the sampled user-item pairs.}
  \label{fig: model}
\end{figure*}

\subsection{Wasserstein dependent GAT}
To effectively propagate variances and overcome the drawbacks of KL divergence mentioned above, we propose Wasserstein dependent Graph ATtention network (W-GAT) to learn more expressive representations for each user and item. 
Graph attention network \cite{velivckovic2017graph} uses a self-attention mechanism to calculate the weight coefficients between nodes and their neighbors.

%e_{ij} = a(\textbf{W}\textbf{h}_i,\textbf{W}\textbf{h}_j)
\begin{equation}\label{GAT att1}
\alpha_{ui}=\frac{\exp(a(\textbf{W}\textbf{h}_u,\textbf{W}\textbf{h}_i))}{\mathbf{\Sigma}_{j\in\mathcal{N}_u}\exp(a(\textbf{W}\textbf{h}_u,\textbf{W}\textbf{h}_j))}
\end{equation}
$\alpha_{ui}$ is the attention weight between user $u$ and item $i$. $\textbf{W}$ is the masked attention weight matirx and $\textbf{h}$ is the point embedding of a node. $a(\cdot)$ is a linear transformation function with the LeakyReLU activation function. 

It is inappropriate to directly extend the method of computing attention for point embeddings to probabilistic embeddings. The variances of users and items are naturally distinct in semantics, so concatenating them arbitrarily will inevitably cause confusion. Besides, many attention mechanisms resort to the inner product to calculate attention score, which is unsuitable for capturing the Wasserstein distance. The inner product lacks a direct relationship with Euclidean distance and Wasserstein distance. Notably, both large and small inner products can lead to significant Wasserstein distances, which complicates the task of distinguishing important items for a given user. This incompatibility of the inner product poses challenges in accurately capturing the relevance and importance of items in the attention process.

Thus, we introduce Wasserstein distance as attention scores to measure relationships between pairs of users and items, defined as follows,

\iffalse
\begin{equation}\label{W-attention 1}
    e_{ui} = -W_2(\textbf{e}_u,\textbf{e}_i)
\end{equation}
\fi

\begin{equation}\label{W-attention}
\alpha_{ui}=\frac{\exp(-W_2(\textbf{e}_u,\textbf{e}_i))}{\mathbf{\Sigma}_{j\in\mathcal{N}_u}\exp(-W_2(\textbf{e}_u,\textbf{e}_j))}
\end{equation}
$\textbf{e}_u$ and $\textbf{e}_i$ represents the Gaussian distributions of user $u$ and item $i$. W-GAT updates node embeddings as follows, 
 
\begin{equation}\label{W-GAT 1}
\begin{aligned}
    & \bm{\mu}_i^{(k+1)} = \Tilde{A}^k \bm{\mu}_i^{(k)},\bm{\mu}_i = \frac{1}{K+1} \Sigma^K_{k=0} \Tilde{A}^k \bm{\mu}_i^{(k)}
    \\
    & \mathbf{\Sigma}_i^{(k+1)} = \Tilde{A}^{2k}\mathbf{\Sigma}_i^{(k)}, \mathbf{\Sigma}_i = \frac{1}{K+1}\Sigma^K_{k=0}\Tilde{A}^{2k}\mathbf{\Sigma}_i^{(k)}
\end{aligned}
\end{equation}
$\Tilde{A}$ is the attention weight matrix masked by the adjacency matrix. To preserve the semantic continuity of user and item variances, we propagate the variances by $\Tilde{A}^2$, which ensures the information of variances comes from the nodes that have the same semantics. Formally, the prediction score of user $u$ and a candidate item $i$ can be formulated by Wasserstein distance as the following,
\begin{equation}\label{W score}
    \hat{y}_{u,i} = -W_2(\textbf{e}_u,\textbf{e}_i)
\end{equation}
We utilize the widely adopted Bayesian Personalized Ranking (BPR) loss \cite{rendle2012bpr} as the primary loss function in our method, which is defined as follows: 
\begin{equation}\label{L_BPR}
    L_{BPR} = \Sigma_{(u,i,j)\in \mathcal{D}}\left(-\log \sigma(\hat{y}_{u,j}-\hat{y}_{u,i})\right),
\end{equation}
where $(u,i,j)$ is a randomly sampled triplet from the training set $\mathcal{D}$. $i$ is an interacted item of user $u$ and item $j$ is a randomly sampled negative item. $\sigma(\cdot)$ denotes the sigmoid function.

\subsection{Wasserstein Dependent Mutual Information}

Many methods aim to maximize the mutual information between observed data and learned representation to ensure the representation retains the most information about the underlying data. The widely used contrastive learning loss\cite{wu2022effectiveness} maximizes the mutual information between positive pairs to enhance performances.

\subsubsection{Limitations of Mutual Information by KL divergence}
Formally, mutual information measured by KL divergence is given as,
\begin{equation}\label{MI by KL}
    I(u,i) = D_{KL}(p_{(u,i)}||p_u\cdot p_i)
\end{equation}
where $u$ and $i$ are different variables and $p_u$ and $p_i$ are the corresponding PDFs. However, in the context of probabilistic embedding, the mutual information computed by KL divergence inherits the limitations of the KL divergence we mentioned in \ref{sec: Gaussian on graph}. 

Firstly, KL divergence-based mutual information estimation is sensitive to minor differences, which means that the mutual information $I(u,i)$, can be significant even when there are only minor differences between the joint probability distribution $p_{(u,i)}$ and the product of marginal distributions $p_u\cdot p_i$. The latter signifies that variables $p_u$ and $p_i$ are nearly independent, implying limited mutual information between them. In the context of collaborative filtering, it may result in significant mutual information between user $u$ and an irrelevant item $i$, further leading to biased user preferences and subsequently a suboptimal performance.

Secondly, a previous study\cite{arjovsky2017wasserstein} has shown that models encounter vanishing gradient problems when dealing with non-overlapping distributions. Since $D_{KL}(p_{(u,i)}||p_u\cdot p_i)$ is undefined or infinite when $p_u\cdot p_i$ is 0 at some point. Thus, it results in an unstable training process when estimating mutual information.

Lastly, the exponential size of sampling is required to approximate the lower bound of $D_{KL}$. It stems from the limitations of KL divergence as a measure of distribution similarity and from the result derived by \cite{mcallester2020formal} in Theorem \ref{Theorem 1}.

\begin{theorem}\label{Theorem 1}
Let p(x) and q(x) be two distributions and R and S be two sets of n samples from p(x) and q(x), respectively. Let $\delta$ be a confident parameter, and let B(R,S,$\delta$) be a real-valued function of the two samples S and R and the confidence parameter $\delta$. We have that, if with probability at least $1-\delta$,
\begin{equation*}
    B(R,S,\delta)\leq D_{KL}(p(x)||q(x))
\end{equation*}
then with probability at least $1-4\delta$, we have
\begin{equation*}
    B(R,S,\delta)\leq \log n
\end{equation*}
\end{theorem}
It can be concluded that any high confidence lower bound on the mutual information by KL divergence requires an exponential sampling size. 

\subsubsection{Wasserstein dependent Mutual Information with Lipschitz Constraint}

Ozair et al. \cite{ozair2019wasserstein} first pointed out that the Wasserstein distance can be used to measure the mutual information between two distributions and proposed $L_{WPC}$.
To address the aforementioned limitations of KL divergence in computing mutual information, we incorporate Wasserstein dependent mutual information into collaborative filtering, as shown below. 
\begin{equation}\label{W-MI}
\begin{aligned}
        I_W(u,i) & \stackrel{\text{def}}{=}W_2(p_{(u,i)},p_up_i) \\
        & = \sup_{f\in \mathcal{L}_{M\times M}} \mathbb{E}_{p_{(u,i)}}[f(u,i)]-\mathbb{E}_{p_up_i}[f(u,i)]
\end{aligned}
\end{equation}
\begin{multline}\label{L_WPC}
       L_{WPC} = \sup_{f\in \mathcal{L}_{M\times M}} \mathbb{E}_{p_{(u,i)}}[f(u,i)]- \\ \mathbb{E}_{p_up_j}\left[\log \Sigma_j \exp f(u,j)\right]
\end{multline}
$\mathcal{L}_{M\times M}$is the set of all Lipschitz constrained function in $M\times M \in \mathbb{R}$. $i$ is a positive sample of user $u$ and $j$ is a randomly sampled negative item. $f$ is the similarity measurement. Empirical results in \cite{ozair2019wasserstein} demonstrate that the $L_{WPC}$ exhibits a more stable training process with fewer samples.

%$L_{WPC}$ is a lower bound on both contrastive predictive coding and the dual Wasserstein dependency measure \textcolor{red}{[cite WDM learning]}.

Since any positive real score function can be used as $f$ \cite{oord2018representation}, we directly utilize the Wasserstein distance to calculate the predicted score, as shown below.
\begin{equation}
\label{f(u,i)}
    f(u,i)=\frac{1}{\tau}\sigma(-W_2(\textbf{e}_u,\textbf{e}_i))
\end{equation}
$\tau$ is the temperature parameter commonly used to control the hardness level in contrastive loss. $\sigma(\cdot)$ is the sigmoid function to maintain the 1-Lipschitz constraint. Finally, the model will calculate the similarity between each user and all items, and recommend the top K most similar items to the user.

In the collaborative filtering scenario, contrastive learning techniques reduce the distance between positive pairs of user-item interactions while increasing the distance between negative pairs. According to \cite{ozair2019wasserstein}, an encoder without a Lipschitz constraint would exaggerate minor differences between the user and negative sample items beyond the actual underlying dissimilarity. The overestimation can lead the model to incorrectly classify some false negative items, which may actually be potentially positive, as true negative items, biasing the learning of accurate user preferences.
%It further hinders the model to distinguish other negative pairs since a noticeable difference is enough to minimize the loss objective.
Instead, a forced Lipschitz encoder will learn distances between pairs bounded by the distances between underlying samples and prevent the model from arbitrarily maximizing the distance of negative pairs.

It is demonstrated that the self-attention mechanism based on the L2 norm satisfies the Lipschitz property\cite{kim2021lipschitz}. Thus, Wasserstein dependent attention score calculated by two L2 distances satisfies the Lipschitz property, according to Equation (\ref{define 2-W distance}) and (\ref{W-attention}). Since the score function in $L_{WPC}$ is also Lipschitz-constrained, we prevent the model from unrestrictedly amplifying the distances between negative samples.

\subsubsection{Differences}
\label{differences}
Our work differs from \cite{fan2023mutual} and other VAE based methods\cite{gan2023viga, ding2021semi, zhang2021wasserstein, yao2020correlated} in several aspects. \textbf{a)} Fan et al. \cite{fan2023mutual} incorrectly derives the relationship between contrastive loss and mutual information by $I(x_a,x_b)=\log\left[\frac{p(x_a|x_b)}{p(x_b)}\right]$, where factually it is $\log\left[\frac{p(x_a|x_b)}{p(x_a)}\right]$. It defines mutual information directly as the Wasserstein distance between x and y, i.e., $I_W(x,y) = -W_2(x,y)$ and also mistakenly states that $I_W(x,y)$ is proportional to $\log\left[\frac{p(x_a|x_b)}{p(x_b)}\right]$. Thus, it contains an incorrect mathematical derivation that leads to flawed reasoning logic. Our paper proposes a properly grounded formulation in collaborative filtering. 
\textbf{b)} Fan et al. \cite{fan2023mutual} employs a sequential attention mechanism that leverages Wasserstein-based mutual information for sequential recommendation. Our proposed Wasserstein dependent attention mechanism framework is uniquely designed to address the challenges of uncertainty modeling and mutual information maximization in collaborative filtering on the bipartite graph.
\textbf{c)} We aim to maximize the mutual information between positive pairs by optimizing $L_{WPC}$ while VAE models\cite{gan2023viga, ding2021semi, zhang2021wasserstein, yao2020correlated} aims to learn a distribution representing all users and items by optimizing the reconstruction loss to fit the ideal distribution. Compared to VAEs, the GCN-based framework can fully leverage the graph neural network to propagate the means and variances through nodes, thereby capturing information from neighborhoods, and subsequently modeling the user-item bipartite graph structure. W-GAT directly represents users and items as Gaussian distributions for similarity computation. Besides, our model sheds the decoder of the VAE model, i.e., there is no need to sample from distributions and use the reparameterization to train.

\subsection{Training Details}
To effectively learn the parameters of our proposed W-GAT, we optimize the following loss function as follows,
\begin{equation}
    \label{Loss}
    L = L_{BPR} + \omega L_{WPC} + \lambda (||\bm{\mu}||_2^2+ ||\mathbf{\Sigma}||_2^2)
\end{equation}
$L_{BPR}$ is defined by equation (\ref{L_BPR}), and we resort to in-batch sample strategy to calculate $L_{WPC}$, where the users in positive pairs consider items in other pairs as negatives. $\omega$ represents a hyper-parameter that controls the balance between the two loss functions, and $\lambda$ controls the $L_2$ regularization strength. 

We attach the pseudocode of W-GAT to facilitate future researchers in reproducing our work.
\renewcommand{\algorithmicrequire}{\textbf{Input:}}
\renewcommand{\algorithmicensure}{\textbf{Output:}}
\begin{algorithm}
\caption{The procedure of W-GAT.}
\label{algorithm:1}
\begin{algorithmic}[1]
\REQUIRE The set of users $\mathcal{U}$; the set of items $\mathcal{I}$; the embedding size $D$; the temperature $\tau$; the loss weight $\omega$; the number of training epochs $Epochs$.

\STATE Construct the user-item interaction graph $\mathcal{G}$ on the training set;
\STATE Initialize Gaussian representations of users and items, $\textbf{e}_{u,i} \sim \mathcal{N}(\mu_{u,i},\Sigma_{u,i})$;
\FOR{$T=1$ to $Epochs$}
\STATE Update the Gaussian representations with W-GAT according to Eq.(\ref{W-GAT 1}).
\STATE Obtain the predicted score $f(u, i)$ according to Eq.(\ref{f(u,i)}).
\STATE Obtain the loss function $L$ according to Eq.(\ref{Loss}).
\STATE Update the parameters to minimize $L$.
\ENDFOR
\ENSURE The overall refined user/item representations $\textbf{e}_u^*$, $\textbf{e}_i^*$.
\end{algorithmic}
\end{algorithm}

\subsection{Complexity Analysis of W-GAT}

\begin{table}[b]
\vspace{-3mm}
	\centering
        \caption{Time complexity of different encoders.}
	\begin{tabular}{c |c}
		\hline
	
		%\cline{2-3}
		\multicolumn{1}{c|}{}
		& \multicolumn{1}{c}{Time Complexity}\\
		\hline
            GCN & $O(L|E|d+L|V|d^2)$\\
		LightGCN & $O(L|E|d)$\\
		GAT & $O(L|E|d+L|V|d^2)$\\
		\textbf{W-GAT} & $O(L|E|d^2)$\\
           
		\hline
	\end{tabular}
\label{Time_complexity}
\end{table}

Since W-GAT uses Wasserstein distance to calculate attention weights, it is necessary to analyze its complexity and compare it with other encoders, like GCN \cite{kipf2016semi}, LightGCN \cite{he2020lightgcn}, and GAT \cite{velivckovic2017graph}. We list the time complexity of different encoders in Table \ref{Time_complexity}. According to \cite{chen2020scalable}, the time complexity of GCN is $O(L|E|d+L|V|d^2)$, where $|V|$ is the number of nodes, $L$ is the number of layers, $d$ is the embedding size, and $|E|$ is the number of edges in the graph.. The time complexity of GAT is comparable to GCN \cite{velivckovic2017graph}, also $O(L|E|d+L|V|d^2)$. Based on the \cite{chen2020scalable} and \cite{he2020lightgcn}, the time complexity of LightGCN is $O(L|E|d)$. Compared to LightGCN, W-GAT adds the computation of attention weights. The weight for each edge is calculated using the Wasserstein distance. Thus, the time complexity of W-GAT is $d$ times that of LightGCN, i.e., $O(L|E|d^2)$. However, since $d$ is much smaller than $d<<|E|$, it is less than GCN or GAT. Even though the time complexity is higher than LightGCN, it remains within an acceptable range.

\begin{table*}[htbp]
	\centering
        \caption{Recommendation performance comparison of all considered baselines. The best performing values are boldfaced.}
	% '+' represent the relative improvement for each baseline.}
	%on BlogCatalog, PPI and Wikipedia dataset
	%\setlength{\tabcolsep}{5.7mm}{
	\begin{tabular}{c |c c |c c |c c}
		\hline
		%\multirow{3}{*}{{\bf Methods}} 
		%\cline{2-13}
		\multicolumn{1}{c|}{}
		& \multicolumn{2}{c|}{M1M}
		& \multicolumn{2}{c|}{DM}
		& \multicolumn{2}{c}{MIND}\\
		\cline{2-7}
		\multicolumn{1}{c|}{}
		& \multicolumn{1}{c}{Recall@20}
		& \multicolumn{1}{c|}{NDCG@20}
		& \multicolumn{1}{c}{Recall@20}
		& \multicolumn{1}{c|}{NDCG@20}
		& \multicolumn{1}{c}{Recall@20}
		& \multicolumn{1}{c}{NDCG@20}\\
		\hline
		MF & 0.2168 & 0.3111 & 0.1495 & 0.0906 & 0.0556 & 0.0390 \\
		NCF & 0.2317 & 0.2386 & 0.3186 & 0.2889 & 0.0671 & 0.0471 \\
            SSM & 0.2103 & 0.2690 & 0.3349 & 0.2523 & 0.0830 & \textbf{0.0629} \\
		GAT+BPR & 0.2315 & 0.3236 & 0.2692 & 0.1948 & 0.0607 & 0.0387 \\
		GAT+InfoNCE & 0.1291 & 0.1318 & 0.3221 & 0.3090 & 0.0684 & 0.0503 \\
		Graph2Gauss & 0.1220 & 0.1709 & 0.2490 & 0.1817 & 0.0325 & 0.0211 \\
		MoG & 0.1415 & 0.1817 & 0.2664 & 0.2189 & 0.0829 & 0.0591 \\
            GER & 0.2414 & 0.3314 & 0.3374 & 0.3082 & 0.0753 & 0.0536 \\
            %GeRec &  &  &  &  &  &\\
		\textbf{W-GAT} & \textbf{0.2514} & \textbf{0.3330} & \textbf{0.3554} & \textbf{0.3420} & \textbf{0.0840} & 0.0595 \\
		\hline
	\end{tabular}
	%}
	
	\label{overall_performance}
\end{table*}

\section{Experiments}
In this section, we conduct extensive experiments on three datasets to answer the following research questions.
\begin{itemize}
\item \textbf{RQ1}: Does our proposed method outperform different methods in collaborative filtering, such as SSM\cite{wu2022effectiveness}, GER\cite{dos2017gaussian} and MoG\cite{oh2018modeling}?
\item  \textbf{RQ2}: Can our method effectively capture uncertain information of the users and items?
\item \textbf{RQ3}: If we replace the component of our model to LightGCN and KL divergence, how do they impact the performance?
\item \textbf{RQ4}: Which component is essential to our model, and how do the hyper-parameters impact the performance?
\end{itemize}

\subsection{Experimental Settings}
\subsubsection{Datasets} We experimented with three public real-world datasets: Movielens-1M \footnote{\url{https://grouplens.org/datasets/movielens/}}, Digital\_Music \footnote{\url{https://nijianmo.github.io/amazon/index.html}} and MIND \footnote{\url{https://msnews.github.io}}. M1M and DM are short for Movielens-1M and Digital\_Music, respectively. M1M dataset is from the MovieLens website. DM dataset is from the public Amazon review dataset. MIND dataset is collected from anonymized behavior logs of the Microsoft News website. We adopt the standard 5-core preprocessing to filter out users or items with less than five interactions, and for MIND, we adopt 10-core. The detailed datasets statistics are summarized in Table \ref{datasets}.

\begin{table}[h]
\centering
\vspace{-2mm}
\caption{Statistics of the three datasets}
%\resizebox{.95\columnwidth}{!}{
\begin{tabular}{l|c|c|c}
    \hline
     & M1M & DM & MIND \\
    \hline
    \#users & 6039 & 16502 & 32643\\
    \#items & 3628 & 11794  & 12108\\
    \#interactions & 666765 & 126587 & 890241\\
    \#Categories & 19 & - & 17\\
    Sparsity & 0.03043 & 0.00065 & 0.00225\\
    \hline
\end{tabular}
\vspace{-2mm}
\label{datasets}
\end{table}

\subsubsection{Baselines} We compare the proposed method with two groups of recommendation baselines. The first group consists of methods based on point embedding, including \textbf{MF}\cite{koren2009matrix}, \textbf{NCF}\cite{he2017neural}, \textbf{GAT}\cite{velivckovic2017graph}, and \textbf{SSM}\cite{wu2022effectiveness}. We employ BPR and InfoNCE loss functions on GAT, followed by +BPR and +InfoNCE. Notably, we didn't include the original LightGCN \cite{he2020lightgcn} in our comparisons, since SSM is the improved version using the InfoNCE loss function. 
The second group of methods which represent users and items as distributions include \textbf{GER}\cite{dos2017gaussian}, \textbf{Graph2Gauss}\cite{bojchevski2017deep}, \textbf{MoG}\cite{oh2018modeling} and \textbf{GeRec}\cite{jiang2020convolutional}\footnote{We tried to reproduce the method by \url{https://github.com/junyji/gaussian-recommender} but failed to achieve comparable performances}. We did not choose VAE-based models as baselines since the frameworks and optimization objectives of such approaches are fundamentally different from ours as mentioned in Section \ref{differences}. We included the most basic contrastive learning method (SSM). Other CL methods were not compared because our focus was solely on the loss function, not on auxiliary tasks of CL such as graph enhancement \cite{liu2023simgcl} or model enhancement \cite{qiu2022contrastive}.

\subsubsection{Evaluation Metrics} We resort to two commonly employed ranking-based metrics to assess the effectiveness of all considered methods: $Recall@K$ and $NDCG@K$. Additionally, we consider all items except those with which the user has interacted in the training set when predicting rankings.
%To elaborate, Recall@$K$ quantifies the average number of items that users engage with, among the top-$K$ candidates. Furthermore, NDCG@$K$ is a precision-centric metric that factors in the predicted position of the ground truth instance.
%In both cases, larger values indicate superior performance. 
Following \cite{he2020lightgcn, wu2022effectiveness}, we fix $K$ at 20 and present the average metrics for all users within the test set.

\subsubsection{Parameter Settings}
We use Adam \cite{kingma2014adam} optimizer and employ Xavier initialization. Embedding size and batch size are fixed to 64 and 2048, respectively. We set the learning rate to $1e-3$ and $L_2$ regularization coefficient $\lambda$ to $1e-5$ for all models. The number of layers for GNNs is assigned to 2. We apply GridSearch to choose the best temperature $\tau$ over $\{0.01,0.02,...,0.1,0.15,0.2,...,0.9\}$ and $L_{WPC}$ weight $\omega$ over $\{0.01,0.05,0.1,0.5,1\}$. All other hyper-parameters are specified according to the suggested settings in the original papers. We repeated experiments of each baseline for 5 times and present the average results. All models are trained on a single NVIDIA Tesla V100 GPU.

\subsection{Overall Performances (RQ1)}

Based on the experimental outlined setup, we have conducted extensive evaluations that substantiate the effectiveness of W-GAT. Experimental results show significant improvements in recommendation metrics from multiple perspectives, validating the efficacy of our method. 
Illustrated in Table \ref{overall_performance}, W-GAT consistently outperforms all the other baselines in terms of two metrics considered on both M1M and DM datasets. Especially on the DM dataset, our method exhibits an impressive improvement of 5.33\% in $Recall@20$ and 10.68\% in $NDCG@20$. On the MIND dataset, W-GAT outperforms the others concerning $Recall@20$ while remaining competitive with the best model on $NDCG@20$. For the other models based on Gaussian embeddings or KL divergence, they fail to achieve a competitive performance on different datasets. Especially for Graph2Gauss, the unstable outcomes show that the KL-based energy loss function is not applicable to recommendation tasks.
\vspace{-3mm}

\begin{table}[t]
\centering
\vspace{-3mm}
\caption{Learned Average variances of users in MIND datasets}
%\resizebox{.95\columnwidth}{!}{
\begin{tabular}{c|c|c|c}
    \hline
     \multicolumn{2}{c|}{1st kind of uncertain users} & \multicolumn{2}{c}{2nd kind of uncertain users} \\
    \hline
    $o_1$ & variance & $o_2$ & variance\\
    \hline
    0.0$\sim$1.0 & 11.2375 & 0.0$\sim$0.2 & 10.5893 \\
    1.0$\sim$1.5 & 10.9985 & 0.2$\sim$0.4 & 11.1015 \\
    1.5$\sim$2.0 & 10.7700 & 0.4$\sim$0.6 & 12.2088 \\
    2.0$\sim$2.5 & 10.5692 & 0.6$\sim$0.8 & 13.7144 \\
    $>$2.5 & 10.5814 & 0.8$\sim$1.0 & 13.2349 \\
    \hline
\end{tabular}
\label{user_uncertainty}
\end{table}

\subsection{Effectiveness of Capturing Uncertainty}
\subsubsection{Capturing User Uncertainty (RQ2)}

To evaluate whether our proposed W-GAT can effectively capture the preferences of users with uncertainty, following \cite{jiang2020convolutional}, we focus on two kinds of users. For the first kind, we aim to observe the uncertainty introduced by the number of interactions. We divide the users into several groups according to the metric $o_1$, where $o_1$ represents users with $10^{o_1}$ interactions. The smaller the $o_1$, the less information of user $u$, resulting in more uncertainty. For the second kind, we focus on the diversity of interests, dividing users by the metric $o_2$, where $o_2$ is defined as $o_2=1-\frac{2}{K*(K-1)}\sum_{i=1}^K\sum_{j=i+1}^K\mathbb{I}_{ij}$, $K$ is the length of recommendation list, $\mathbb{I}_{ij}$ is an indicator function whose value is 1 when item $i$ and item $j$ have the same category and 0 otherwise. Thus, a large $o_2$ indicates more preference diversity, meaning more uncertainty. We display the modulus length of the variance in the group, and the larger the value, the more the uncertainty.

Table \ref{user_uncertainty} presents the results for MIND users. For the first kind of users, our model assigns larger variances to the users with fewer interactions. Users with more than $10^{2.5}$ interactions have slightly larger variances, representing a broader range of interests and increased uncertainty. As for the second group, users with recommendations across more categories have higher variances, indicating extensive interests and more uncertainty. The outcome shows that our model is capable of capturing user uncertainty induced by a lack of behavior information and a wide range of interests.

\begin{figure}[t]
\centering
\includegraphics[width=\linewidth]{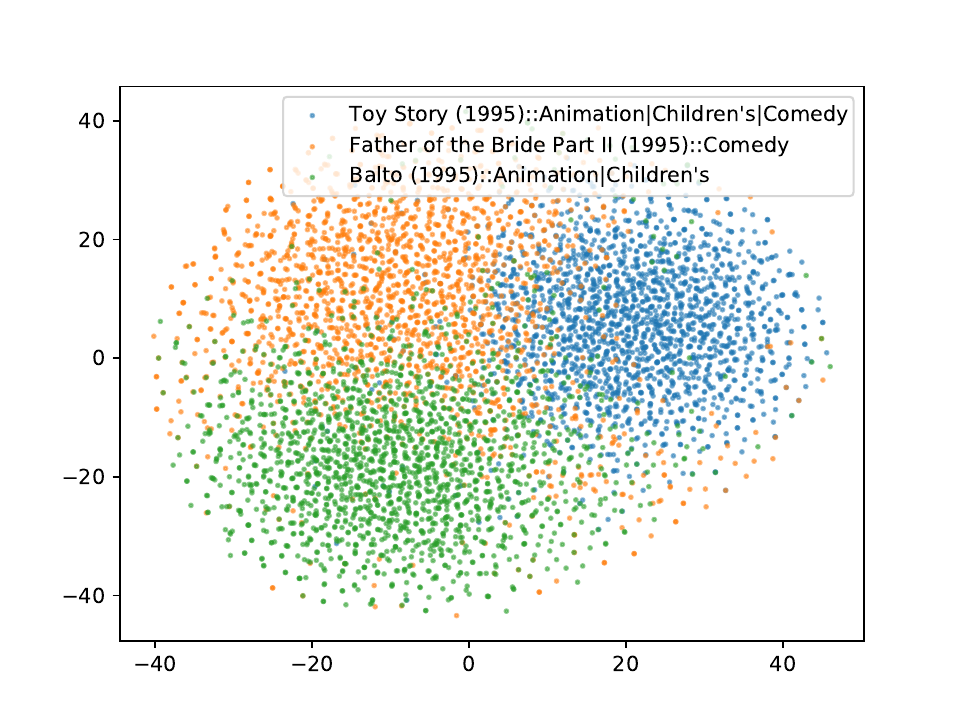}

\centering
\caption{The visualization of Gaussian embeddings of three different movies from M1M}
\label{item uncertainty}
\end{figure}

\subsubsection{Capturing Item Uncertainty (RQ2)}

To verify the effectiveness of the method to model uncertainty information in items, we conducted two following experiments. 
Firstly, we calculate the average variance for items with different numbers of labels, and the results are shown in Table \ref{Average_item_variances}. It is evident that items with only one label have the highest variance, indicating the highest level of uncertainty, while those with three or more labels have the smallest average variance, suggesting lower uncertainty.
Secondly, we visualize the Gaussian embeddings of three movies from MIM shown in Figure \ref{item uncertainty} by T-SNE \cite{van2008visualizing}. Specifically, we selected Toy Story (1995), Father of the Bride Part II (1995), and Balto (1995) as the visualized movies, and they are categorized as Animation|Children's|Comedy, Comedy, and Animation|Children's, respectively. The 1000 different colored points in Figure \ref{item uncertainty} are sampled from different Gaussian embeddings. It's clear that the three distributions intersect with each other, representing the overlap in the classification of the three movies. Moreover, the orange sampling points are the most dispersed while the blue points are the most concentrated, meaning Father of the Bride Part II (1995) has the largest variance and Toy Story (1995) has the smallest. It indicates that the more detailed the classification of the movie is, the more information it contains, the smaller its uncertainty is, and the smaller the variance of the Gaussian embedding learned by the model is. 

Overall, our method could effectively capture the uncertainty from both the user's and the item's perspective, modeling the range of interests of two different groups of users and discovering uncertain information with categories of items, further enhancing recommendation accuracy and robustness.

\begin{table}[h]
	\centering
        \caption{Average variances of items with different numbers of labels}
	\begin{tabular}{c |c}
		\hline
		%\cline{2-3}
		\multicolumn{1}{c|}{Number of Labels}
		& \multicolumn{1}{c}{Item Variance}\\
		\hline
		1 & 12.555668 \\
		2 & 12.547083 \\
            3+ & 12.409182 \\
           
		\hline
	\end{tabular}
 
\label{Average_item_variances}
\end{table}

\begin{figure}[t]
\centering
\subfigure{
\begin{minipage}[t]{0.49\linewidth}
\centering
\includegraphics[width=\linewidth]{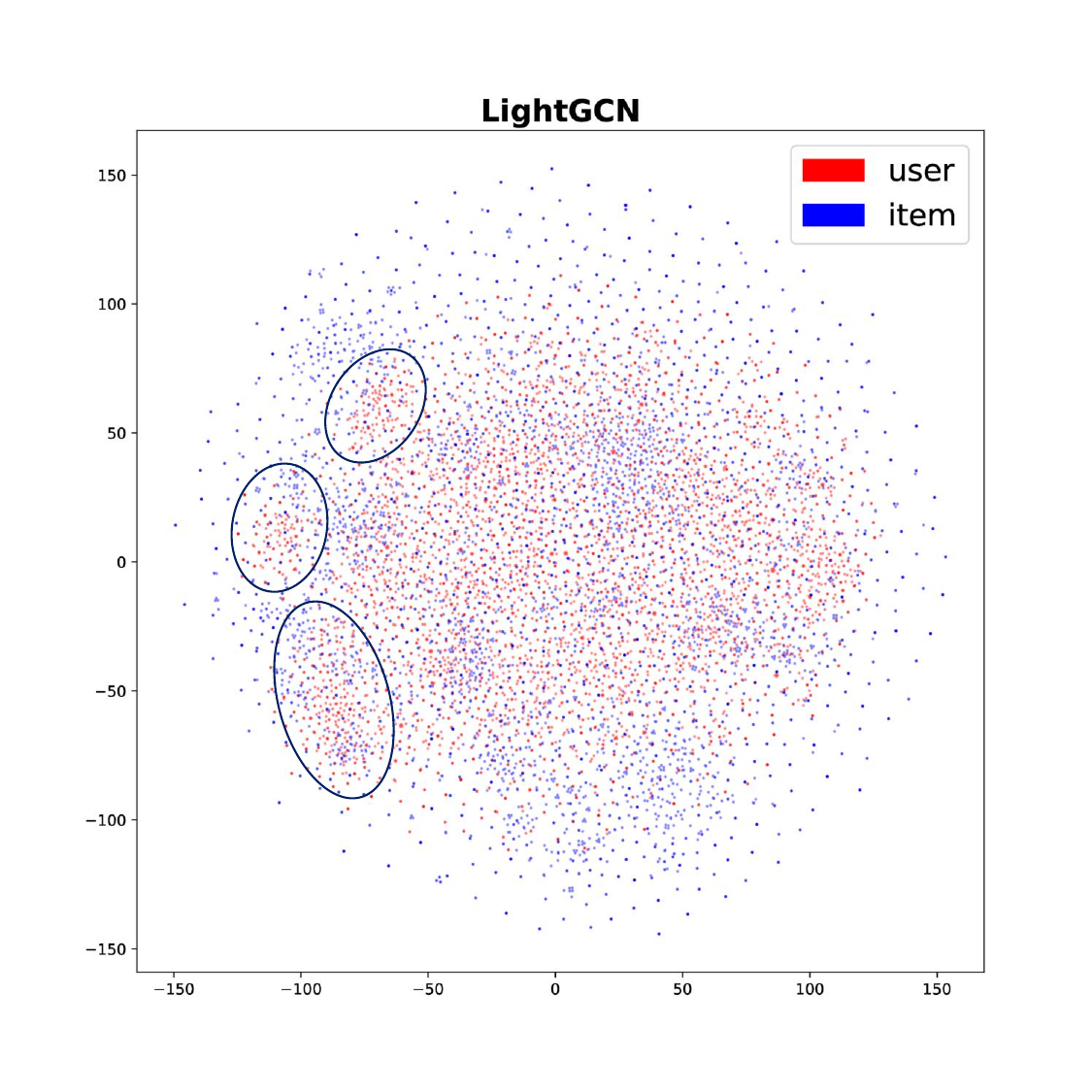}
%\caption{fig1}
\end{minipage}%
}%
\subfigure{
\begin{minipage}[t]{0.49\linewidth}
\centering
\includegraphics[width=\linewidth]{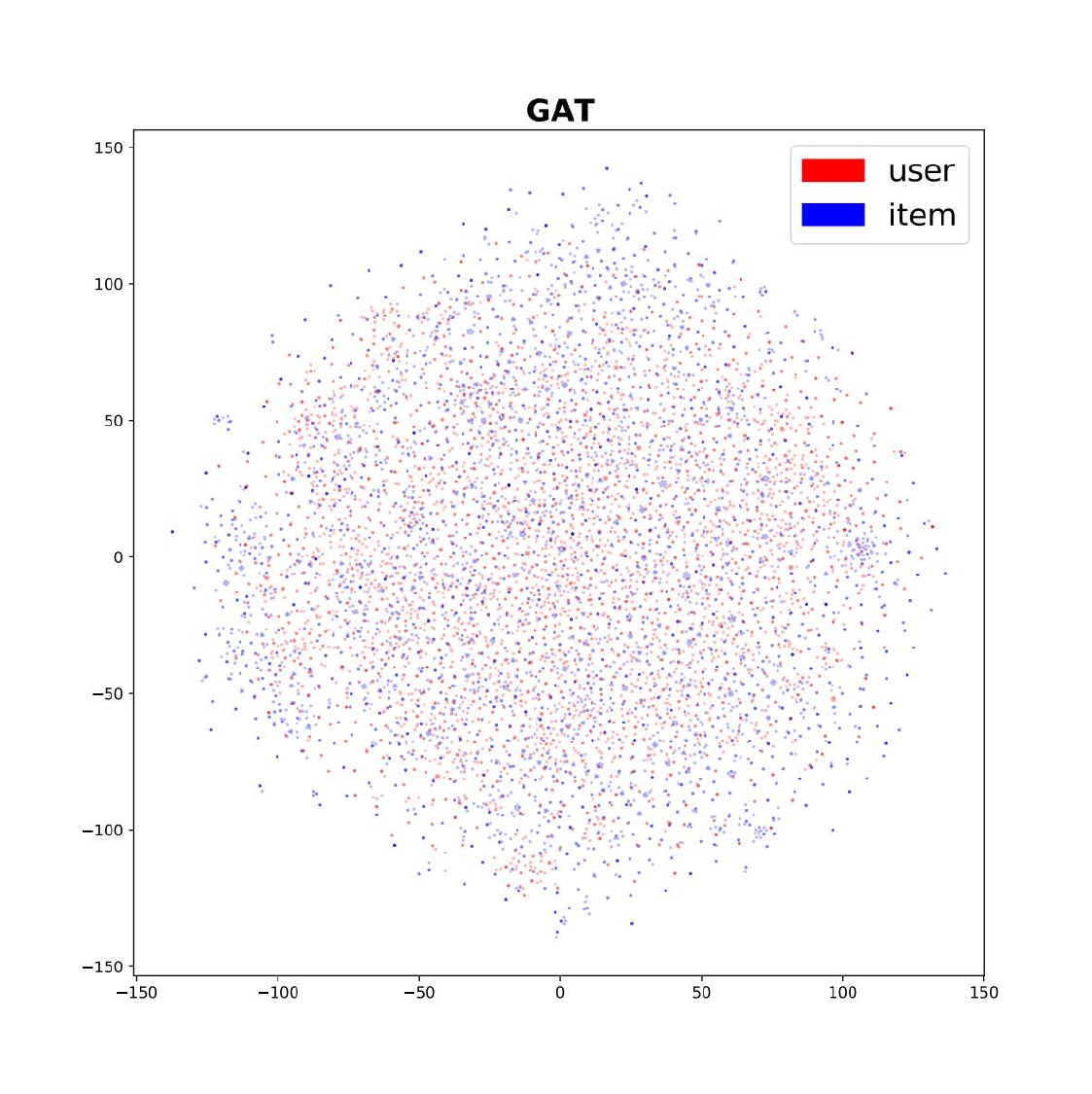}
%\caption{fig2}
\end{minipage}%
}%
\centering
\vspace{-2mm}
\caption{Visualization of mean embedding learned by LightGCN and GAT on the M1M dataset.}
\label{visual}
\end{figure} 

\subsection{Study of W-GAT}
\subsubsection{Limitations of LightGCN (RQ3)}
\label{subsubsec: limitation of LighGCN}
As mentioned in the previous section, LightGCN faces the challenge of unconditional propagation of variances, confusing the semantic interpretation of users and items. We conducted the following experiments to substantiate the suboptimal recommendation performance of LightGCN when propagating variance. We visualized the embeddings of two GNN encoders incorporating the BPR loss function on the M1M dataset. Figure \ref{visual} illustrates their representations by visualizing Gaussian distributions learned by the model on the two-dimensional representation by T-SNE \cite{van2008visualizing}. Notably, only the mean values of the Gaussian distribution are used for better visualization. Observations reveal that unconstrained propagation of variances in LightGCN leads to clustering phenomena among users and items, depicted in Figure \ref{visual}, validating the confusion of two semantics stated in Section \ref{sec: Gaussian on graph}. It causes the model to recommend the closest item to a cluster of users, resulting in suboptimal performance in Table \ref{LightGCN_GAT}. The experimental settings in Table \ref{LightGCN_GAT} are the same as those in Table \ref{overall_performance}. It specifically demonstrates the performance when using LightGCN and GAT to encode Gaussian distributions and training only with BPR loss.

\begin{table}[h]
	\centering
        \caption{Performances of two encoders on the M1M dataset.}
	\begin{tabular}{c |c c}
		\hline
	
		%\cline{2-3}
		\multicolumn{1}{c|}{}
		& \multicolumn{1}{c}{Recall@20}
		& \multicolumn{1}{c}{NDCG@20}\\
		\hline
		LightGCN & 0.2326 & 0.3128  \\
		GAT & 0.2412 & 0.3264  \\
           
		\hline
	\end{tabular}
 
\vspace{-3mm}	
\label{LightGCN_GAT}
\end{table}

\begin{figure*}[tphb]
\centering
\subfigure{
\begin{minipage}[t]{0.33\linewidth}
\centering
\includegraphics[width=\linewidth]{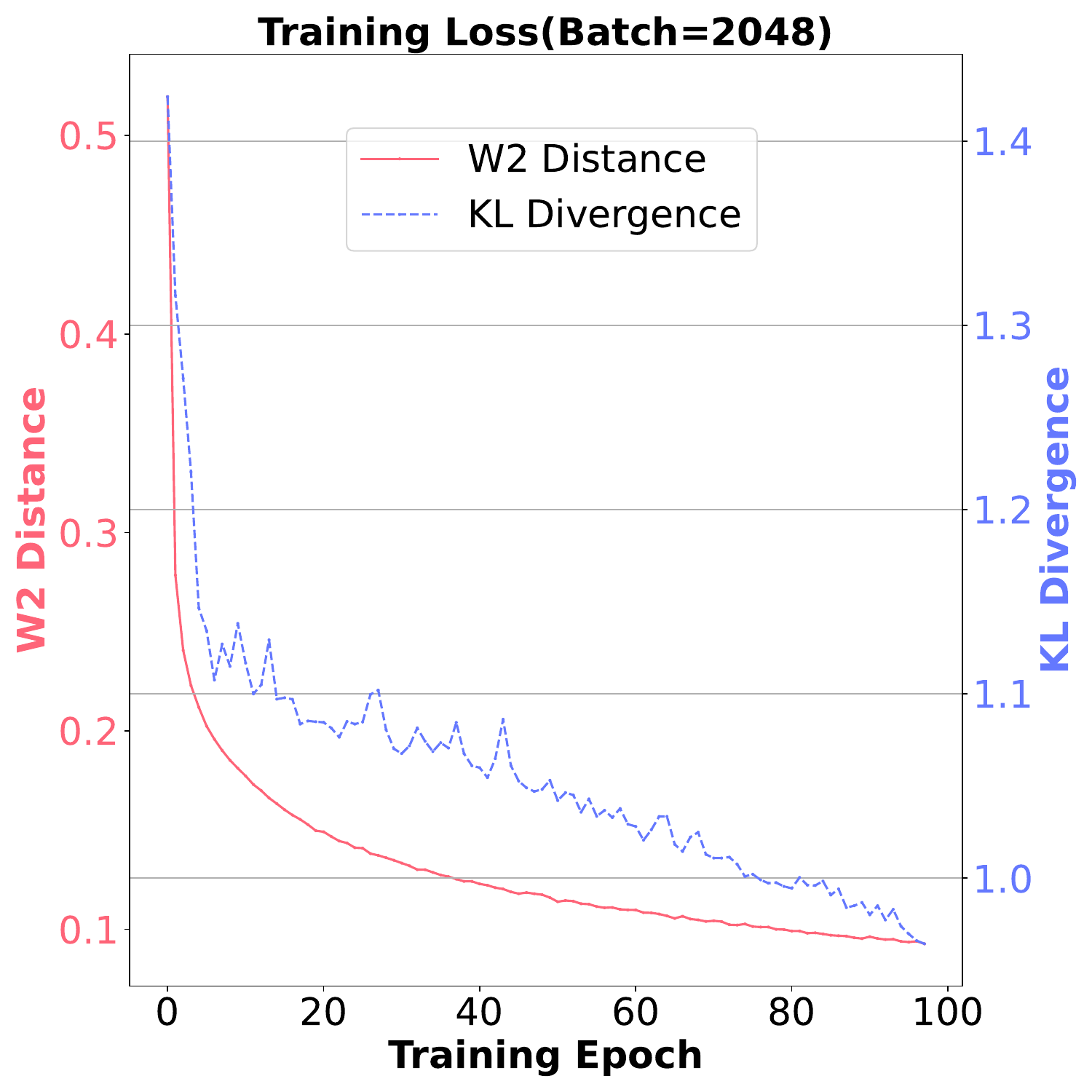}
%\caption{fig1}
\end{minipage}%
}%
\subfigure{
\begin{minipage}[t]{0.33\linewidth}
\centering
\includegraphics[width=\linewidth]{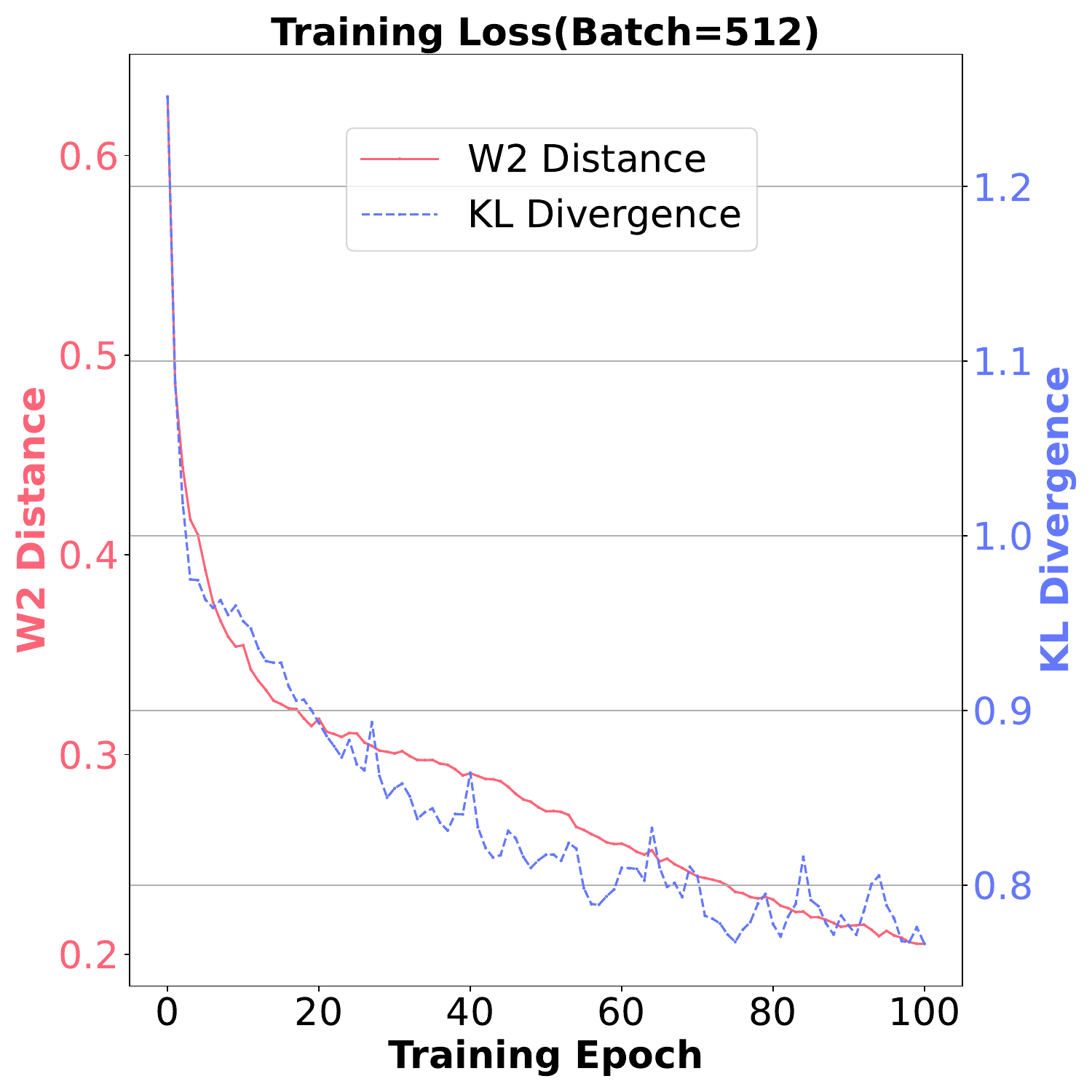}
%\caption{fig2}
\end{minipage}%
}%
\subfigure{
\begin{minipage}[t]{0.33\linewidth}
\centering
\includegraphics[width=\linewidth]{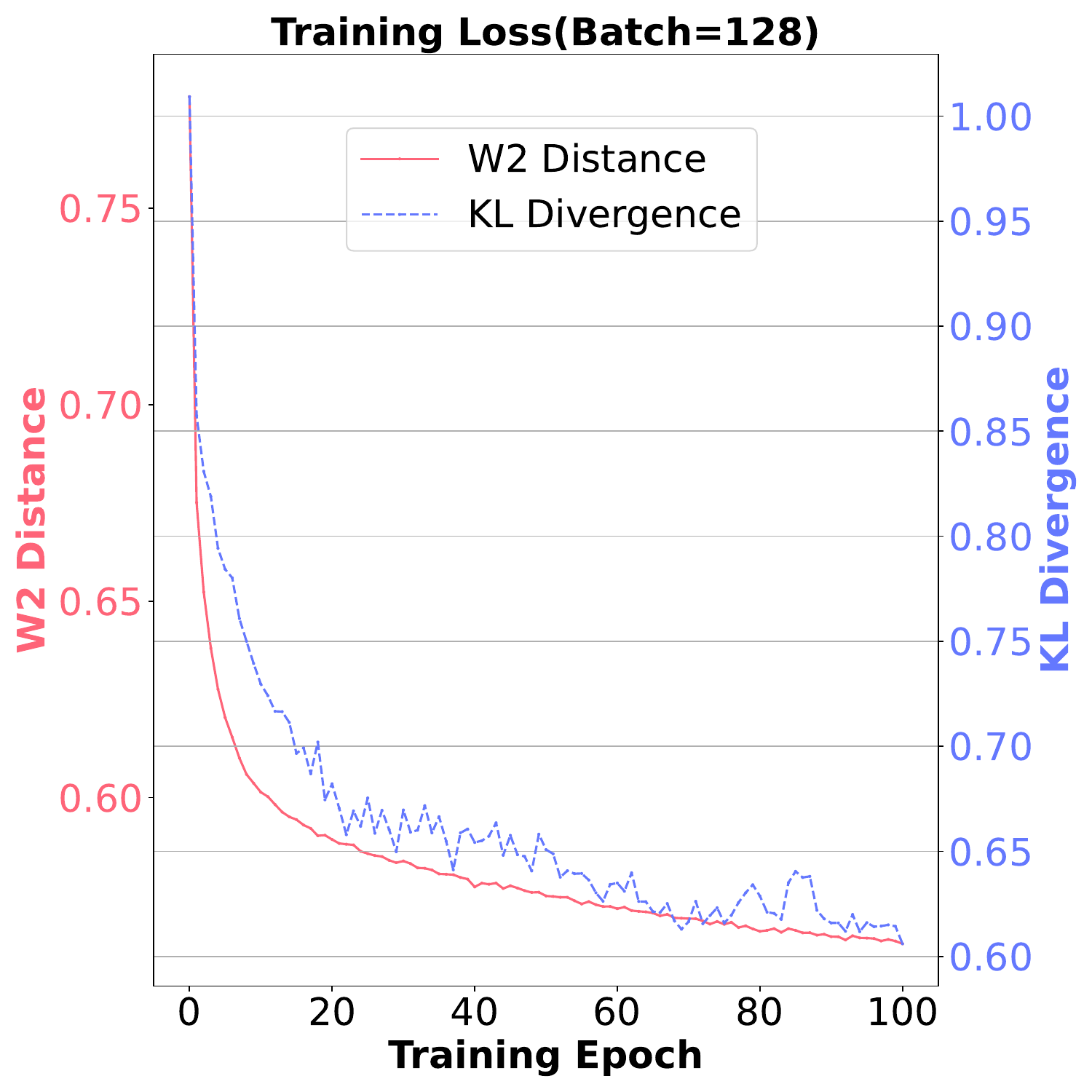}
%\caption{fig2}
\end{minipage}%
}%
\centering
\caption{The values of loss function with two different measurement under different batch sizes.}
\label{fig: limitation of KL}
\end{figure*}

\subsubsection{Limitations of Mutual Information by KL Divergence (RQ3)}
\label{subsubsec: limitation of MI by KL}
To validate the theoretical analysis of the limitations of mutual information measured by KL divergence, we conducted the following experiment. We replaced Wasserstein distance in W-GAT to KL Divergence, and attempted to compare the performance of models using KL divergence and Wasserstein distance. However, the model using KL divergence did not perform as comparable as expected. We speculate that this is due to the gradient vanishing and training instability caused by KL divergence, as mentioned in Section \ref{sec: Gaussian on graph}. Additionally, directly comparing performances does not effectively highlight the reasons behind the suboptimal performance. Therefore, we recorded the trends of loss function under different batch sizes during the training process, depicted in Figure \ref{fig: limitation of KL}. Notably, we only recorded the first 100 epochs because the method based on KL divergence takes hundreds of epochs to converge. Regardless of the batch size, the loss value of the method based on KL divergence experiences a clear oscillating downward trend, while W-GAT shows a relatively steady downward trend, further proving that KL divergence leads to an unstable training process. 
As the batch size decreases, the loss values of the KL Divergence-based method show more pronounced fluctuations, while W-GAT shows weak oscillations under a smaller batch. This phenomenon verifies the theoretical analysis that states KL divergence-based mutual information requires more samples, otherwise, it cannot approximate the lower bound with high confidence.

\subsubsection{Ablation Study (RQ4)}

We conduct ablation studies on W-GAT to understand the impact of each component on its performance. Specifically, we investigate two variants: (1) Remove the auxiliary task $L_{WPC}$ and only employ $L_{BPR}$ to train our model. (2) Use the attention weight matrix $\tilde{A}$ to propagate the variance on the user-item graph instead of $\tilde{A}^2$.

\begin{figure}[htbp]
\centering

\subfigure{
\begin{minipage}[t]{0.5\linewidth}
\centering
\includegraphics[width=\linewidth]{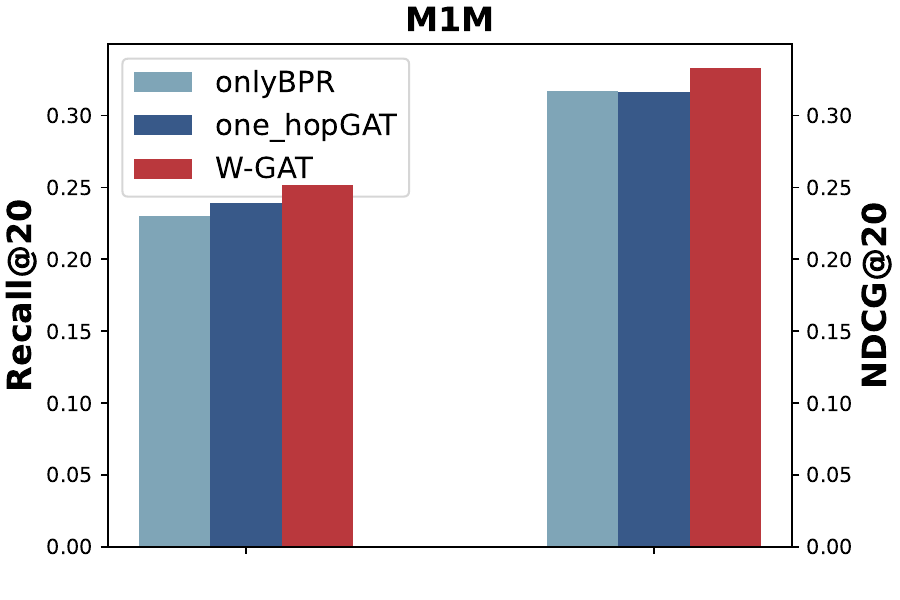}
%\caption{fig1}
\end{minipage}%
}%
\subfigure{
\begin{minipage}[t]{0.5\linewidth}
\centering
\includegraphics[width=\linewidth]{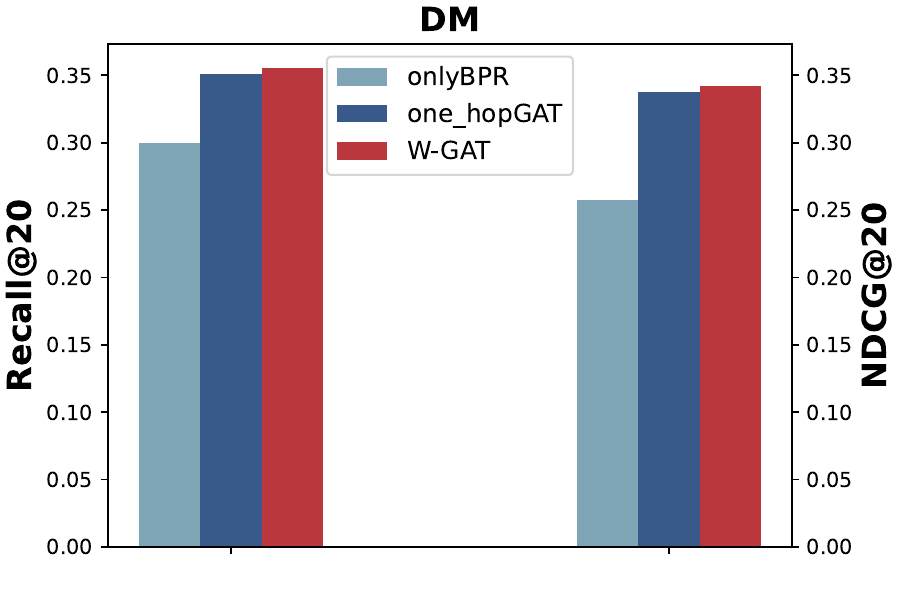}
%\caption{fig2}
\end{minipage}%
}%
\centering
\caption{Ablation study on M1M and DM datasets.}
\label{ablation}
\end{figure}

As shown in Figure \ref{ablation}, any variant is slightly worse than W-GAT on both datasets. Particularly, when removing the auxiliary loss $L_{WPC}$, a more significant drop in performance is evident, indicating that $L_{WPC}$ plays a crucial role in improving the quality of learned embeddings, consequently enhancing recommendation performance. In addition, using $\tilde{A}^2$ to propagate the variance on the user-item interaction graph can obtain better performances. It suggests that variances are better suited for propagating among nodes with the same semantics than propagating to neighbors.

\subsubsection{Parameter Sensitivity Analysis (RQ4)}

\begin{figure}[htbp]
  \centering
  \includegraphics[width=\linewidth]{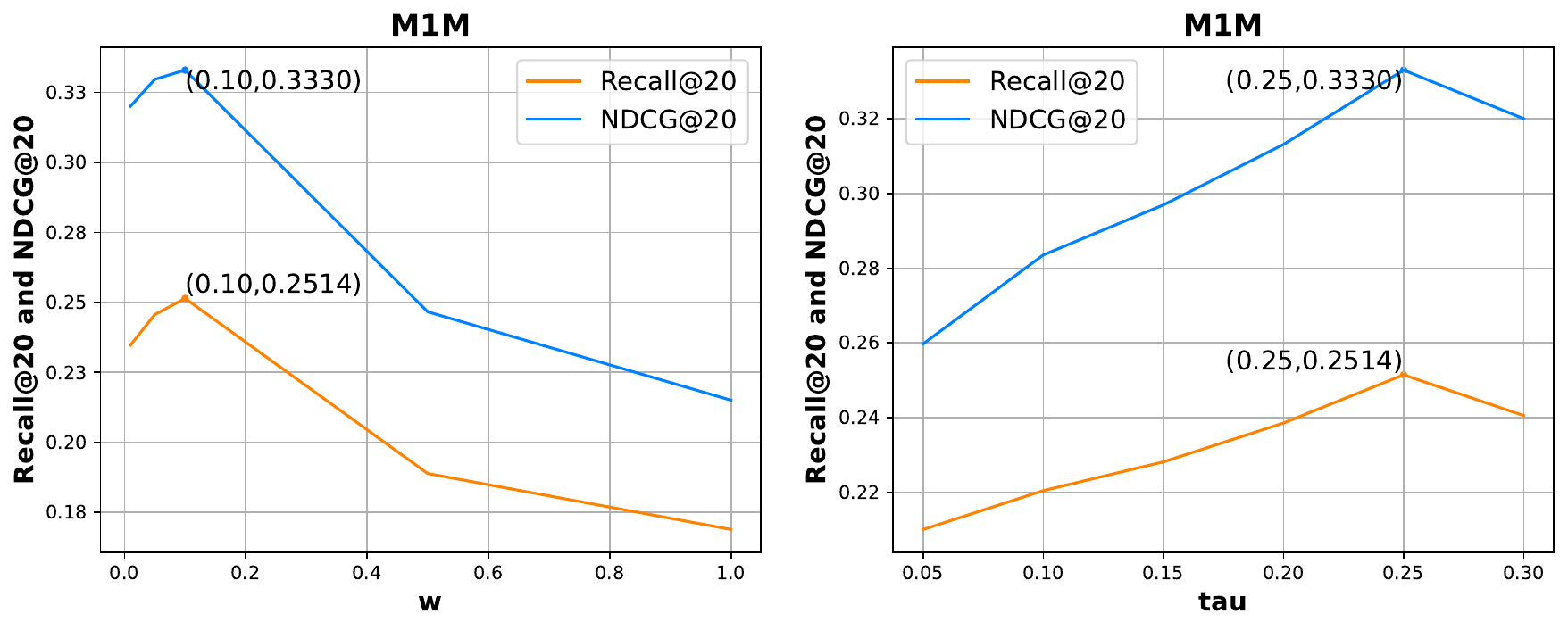}
  \caption{Recall@20 and NDCG@20 of W-GAT obtained with different parameters $\omega$ and $\tau$.}
  \label{para_sen}
\end{figure}

Figure \ref{para_sen} shows the performance change of W-GAT under different loss weights $\omega$ and temperature $\tau$ on the M1M dataset. Notably, we only conducted this experiment on the M1M dataset because the figures on the other two datasets share the same trends. Both $Recall@20$ and $NDCG@20$ exhibit a similar trend when adjusting $\omega$ and $\tau$ within a certain range. In particular, the recommendation performance improves and then drops when the loss weight $\omega$ increases from 0.01 to 1. This behavior is attributed to excessively large $\omega$ overemphasizing the auxiliary tasks, which in turn reduces the weight assigned to the recommendation tasks, ultimately leading to suboptimal performance. In addition, the recommendation performance also follows an initial improvement and subsequent decline with the increase of $\tau$ across the range $\{0.05,...,0.3\}$. In this context, $\tau$ controls the ability to mine negative samples, and better recommendation performance can be achieved by tuning $\tau$ within a proper range. The optimal value of $\omega$ and $\tau$ in the M1M dataset is 0.1 and 0.25, respectively. 

\section{Conclusion}
We proposed W-GAT, a Wasserstein dependent graph attention network for Collaborative Filtering. To capture uncertain information, W-GAT represents users and items as Gaussian distributions and applies the graph attention mechanism to effectively propagate the variance information and calculate attention scores via Wasserstein distance to overcome the limitations of LightGCN and KL divergence. Moreover, in order to maximize the mutual information between users and their interacted items, we incorporate Wasserstein Dependent Measurement into our model and constrain the encoder with Lipschitz property. We conducted extensive empirical studies on three public datasets to demonstrate our model outperforms several common methods and existing Gaussian based strategies. Further analysis provides insights into how W-GAT captures uncertainty for user interests and categories of items. Additionally, we analyzed the importance of each component in our model by visualizing embeddings and conducting an ablation study. 
%Parameter analysis shows the robustness of hyper-parameters for the overall performance. 

In the future, more accurate ways to model the variances of users and items can be explored, and probabilistic representations are also worth exploring in terms of enhancing the diversity of recommendations. In addition, we can also apply the idea of probabilistic representation to hypergraph recommendation systems, a field with great research prospects.

%\section*{Acknowledgments}
%This should be a simple paragraph before the References to thank those individuals and institutions who have supported your work on this article.

\bibliographystyle{IEEEtran}
\bibliography{IEEEexample}

\begin{thebibliography}{10}
\providecommand{\url}[1]{#1}
\csname url@samestyle\endcsname
\providecommand{\newblock}{\relax}
\providecommand{\bibinfo}[2]{#2}
\providecommand{\BIBentrySTDinterwordspacing}{\spaceskip=0pt\relax}
\providecommand{\BIBentryALTinterwordstretchfactor}{4}
\providecommand{\BIBentryALTinterwordspacing}{\spaceskip=\fontdimen2\font plus
\BIBentryALTinterwordstretchfactor\fontdimen3\font minus \fontdimen4\font\relax}
\providecommand{\BIBforeignlanguage}[2]{{%
\expandafter\ifx\csname l@#1\endcsname\relax
\typeout{** WARNING: IEEEtran.bst: No hyphenation pattern has been}%
\typeout{** loaded for the language `#1'. Using the pattern for}%
\typeout{** the default language instead.}%
\else
\language=\csname l@#1\endcsname
\fi
#2}}
\providecommand{\BIBdecl}{\relax}
\BIBdecl

\bibitem{koren2009matrix}
Y.~Koren, R.~Bell, and C.~Volinsky, ``Matrix factorization techniques for recommender systems,'' \emph{Computer}, vol.~42, no.~8, pp. 30--37, 2009.

\bibitem{he2017neural}
X.~He, L.~Liao, H.~Zhang, L.~Nie, X.~Hu, and T.-S. Chua, ``Neural collaborative filtering,'' in \emph{Proceedings of the 26th international conference on world wide web}, 2017, pp. 173--182.

\bibitem{he2020lightgcn}
X.~He, K.~Deng, X.~Wang, Y.~Li, Y.~Zhang, and M.~Wang, ``Lightgcn: Simplifying and powering graph convolution network for recommendation,'' in \emph{Proceedings of the 43rd International ACM SIGIR conference on research and development in Information Retrieval}, 2020, pp. 639--648.

\bibitem{mnih2007probabilistic}
A.~Mnih and R.~R. Salakhutdinov, ``Probabilistic matrix factorization,'' \emph{Advances in neural information processing systems}, vol.~20, 2007.

\bibitem{rendle2012bpr}
S.~Rendle, C.~Freudenthaler, Z.~Gantner, and L.~Schmidt-Thieme, ``Bpr: Bayesian personalized ranking from implicit feedback,'' \emph{arXiv preprint arXiv:1205.2618}, 2012.

\bibitem{bojchevski2017deep}
A.~Bojchevski and S.~G{\"u}nnemann, ``Deep gaussian embedding of graphs: Unsupervised inductive learning via ranking,'' \emph{arXiv preprint arXiv:1707.03815}, 2017.

\bibitem{fan2022sequential}
Z.~Fan, Z.~Liu, Y.~Wang, A.~Wang, Z.~Nazari, L.~Zheng, H.~Peng, and P.~S. Yu, ``Sequential recommendation via stochastic self-attention,'' in \emph{Proceedings of the ACM Web Conference 2022}, 2022, pp. 2036--2047.

\bibitem{oord2018representation}
A.~v.~d. Oord, Y.~Li, and O.~Vinyals, ``Representation learning with contrastive predictive coding,'' \emph{arXiv preprint arXiv:1807.03748}, 2018.

\bibitem{wu2022effectiveness}
J.~Wu, X.~Wang, X.~Gao, J.~Chen, H.~Fu, T.~Qiu, and X.~He, ``On the effectiveness of sampled softmax loss for item recommendation,'' \emph{arXiv preprint arXiv:2201.02327}, 2022.

\bibitem{fan2023mutual}
Z.~Fan, Z.~Liu, H.~Peng, and P.~S. Yu, ``Mutual wasserstein discrepancy minimization for sequential recommendation,'' \emph{arXiv preprint arXiv:2301.12197}, 2023.

\bibitem{mcallester2020formal}
D.~McAllester and K.~Stratos, ``Formal limitations on the measurement of mutual information,'' in \emph{International Conference on Artificial Intelligence and Statistics}.\hskip 1em plus 0.5em minus 0.4em\relax PMLR, 2020, pp. 875--884.

\bibitem{ozair2019wasserstein}
S.~Ozair, C.~Lynch, Y.~Bengio, A.~Van~den Oord, S.~Levine, and P.~Sermanet, ``Wasserstein dependency measure for representation learning,'' \emph{Advances in Neural Information Processing Systems}, vol.~32, 2019.

\bibitem{xue2017deep}
H.-J. Xue, X.~Dai, J.~Zhang, S.~Huang, and J.~Chen, ``Deep matrix factorization models for recommender systems.'' in \emph{IJCAI}, vol.~17.\hskip 1em plus 0.5em minus 0.4em\relax Melbourne, Australia, 2017, pp. 3203--3209.

\bibitem{wang2019neural}
X.~Wang, X.~He, M.~Wang, F.~Feng, and T.-S. Chua, ``Neural graph collaborative filtering,'' in \emph{Proceedings of the 42nd international ACM SIGIR conference on Research and development in Information Retrieval}, 2019, pp. 165--174.

\bibitem{velivckovic2017graph}
P.~Veli{\v{c}}kovi{\'c}, G.~Cucurull, A.~Casanova, A.~Romero, P.~Lio, and Y.~Bengio, ``Graph attention networks,'' \emph{arXiv preprint arXiv:1710.10903}, 2017.

\bibitem{vilnis2014word}
L.~Vilnis and A.~McCallum, ``Word representations via gaussian embedding,'' \emph{arXiv preprint arXiv:1412.6623}, 2014.

\bibitem{he2015learning}
S.~He, K.~Liu, G.~Ji, and J.~Zhao, ``Learning to represent knowledge graphs with gaussian embedding,'' in \emph{Proceedings of the 24th ACM international on conference on information and knowledge management}, 2015, pp. 623--632.

\bibitem{ma2020probabilistic}
C.~Ma, L.~Ma, Y.~Zhang, R.~Tang, X.~Liu, and M.~Coates, ``Probabilistic metric learning with adaptive margin for top-k recommendation,'' in \emph{Proceedings of the 26th ACM SIGKDD International Conference on knowledge discovery \& data mining}, 2020, pp. 1036--1044.

\bibitem{dos2017gaussian}
L.~Dos~Santos, B.~Piwowarski, and P.~Gallinari, ``Gaussian embeddings for collaborative filtering,'' in \emph{Proceedings of the 40th international ACM SIGIR conference on research and development in information retrieval}, 2017, pp. 1065--1068.

\bibitem{jiang2020convolutional}
J.~Jiang, D.~Yang, Y.~Xiao, and C.~Shen, ``Convolutional gaussian embeddings for personalized recommendation with uncertainty,'' \emph{arXiv preprint arXiv:2006.10932}, 2020.

\bibitem{gan2023viga}
M.~Gan and H.~Zhang, ``Viga: A variational graph autoencoder model to infer user interest representations for recommendation,'' \emph{Information Sciences}, vol. 640, p. 119039, 2023.

\bibitem{ding2021semi}
Y.~Ding, Y.~Shi, B.~Chen, C.~Lin, H.~Lu, J.~Li, R.~Tang, and D.~Wang, ``Semi-deterministic and contrastive variational graph autoencoder for recommendation,'' in \emph{Proceedings of the 30th ACM International Conference on Information \& Knowledge Management}, 2021, pp. 382--391.

\bibitem{zhang2021wasserstein}
X.~Zhang, J.~Zhong, and K.~Liu, ``Wasserstein autoencoders for collaborative filtering,'' \emph{Neural Computing and Applications}, vol.~33, no.~7, pp. 2793--2802, 2021.

\bibitem{yao2020correlated}
L.~Yao, J.~Zhong, X.~Zhang, and L.~Luo, ``Correlated wasserstein autoencoder for implicit data recommendation,'' in \emph{2020 IEEE/WIC/ACM International Joint Conference on Web Intelligence and Intelligent Agent Technology (WI-IAT)}.\hskip 1em plus 0.5em minus 0.4em\relax IEEE, 2020, pp. 417--422.

\bibitem{clement2008elementary}
P.~Clement and W.~Desch, ``An elementary proof of the triangle inequality for the wasserstein metric,'' \emph{Proceedings of the American Mathematical Society}, vol. 136, no.~1, pp. 333--339, 2008.

\bibitem{arjovsky2017wasserstein}
M.~Arjovsky, S.~Chintala, and L.~Bottou, ``Wasserstein generative adversarial networks,'' in \emph{International conference on machine learning}.\hskip 1em plus 0.5em minus 0.4em\relax PMLR, 2017, pp. 214--223.

\bibitem{kim2021lipschitz}
H.~Kim, G.~Papamakarios, and A.~Mnih, ``The lipschitz constant of self-attention,'' in \emph{International Conference on Machine Learning}.\hskip 1em plus 0.5em minus 0.4em\relax PMLR, 2021, pp. 5562--5571.

\bibitem{kipf2016semi}
T.~N. Kipf and M.~Welling, ``Semi-supervised classification with graph convolutional networks,'' \emph{arXiv preprint arXiv:1609.02907}, 2016.

\bibitem{chen2020scalable}
M.~Chen, Z.~Wei, B.~Ding, Y.~Li, Y.~Yuan, X.~Du, and J.-R. Wen, ``Scalable graph neural networks via bidirectional propagation,'' \emph{Advances in neural information processing systems}, vol.~33, pp. 14\,556--14\,566, 2020.

\bibitem{oh2018modeling}
S.~J. Oh, K.~Murphy, J.~Pan, J.~Roth, F.~Schroff, and A.~Gallagher, ``Modeling uncertainty with hedged instance embedding,'' \emph{arXiv preprint arXiv:1810.00319}, 2018.

\bibitem{liu2023simgcl}
C.~Liu, C.~Yu, N.~Gui, Z.~Yu, and S.~Deng, ``Simgcl: graph contrastive learning by finding homophily in heterophily,'' \emph{Knowledge and Information Systems}, pp. 1--26, 2023.

\bibitem{qiu2022contrastive}
R.~Qiu, Z.~Huang, H.~Yin, and Z.~Wang, ``Contrastive learning for representation degeneration problem in sequential recommendation,'' in \emph{Proceedings of the fifteenth ACM international conference on web search and data mining}, 2022, pp. 813--823.

\bibitem{kingma2014adam}
D.~P. Kingma and J.~Ba, ``Adam: A method for stochastic optimization,'' \emph{arXiv preprint arXiv:1412.6980}, 2014.

\bibitem{van2008visualizing}
L.~Van~der Maaten and G.~Hinton, ``Visualizing data using t-sne.'' \emph{Journal of machine learning research}, vol.~9, no.~11, 2008.

\end{thebibliography}

\vspace{-30pt}
\begin{IEEEbiography}[{\includegraphics[width=1in,height=1.25in,clip,keepaspectratio]{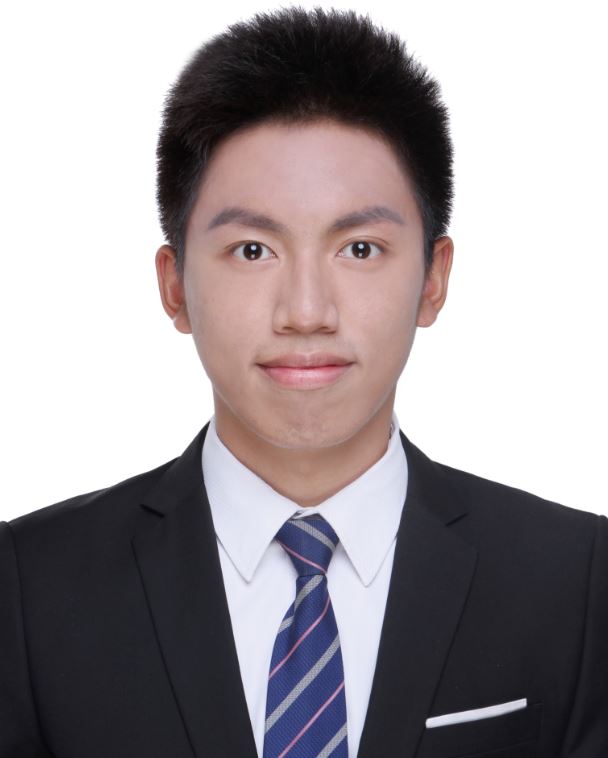}}]{Haoxuan Li}
received a BSc degree from Beihang University in 2018. He is currently studying for a master's degree in School of Computer Science and Engineering and with the Engineering Research Center of Advanced Computer Application Technology, Ministry of Education, Beihang University. His main research interests cover educational data mining, uncertainty, recommendation.
\end{IEEEbiography}
\vspace{-10 mm}
\begin{IEEEbiography}[{\includegraphics[width=1in,height=1.25in,clip,keepaspectratio]{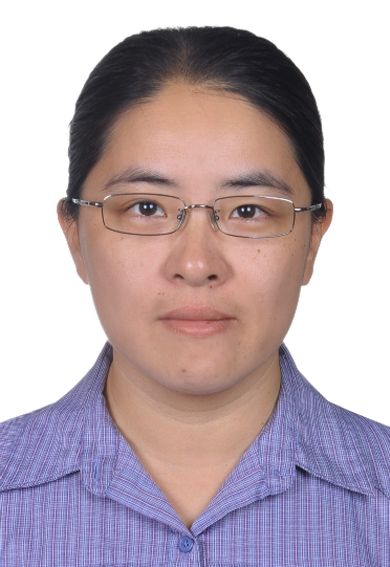}}]{Yuanxin Ouyang}
is a professor in School of Computer Science and Engineering, Beihang University, China. She received PhD and BSc degrees from Beihang University in 2005, 1997, respectively. Her area of research covers recommender system, educational data mining, social networks and service computing.
\end{IEEEbiography}
\vspace{-10 mm}
\begin{IEEEbiography}[{\includegraphics[width=1in,height=1.25in,clip,keepaspectratio]{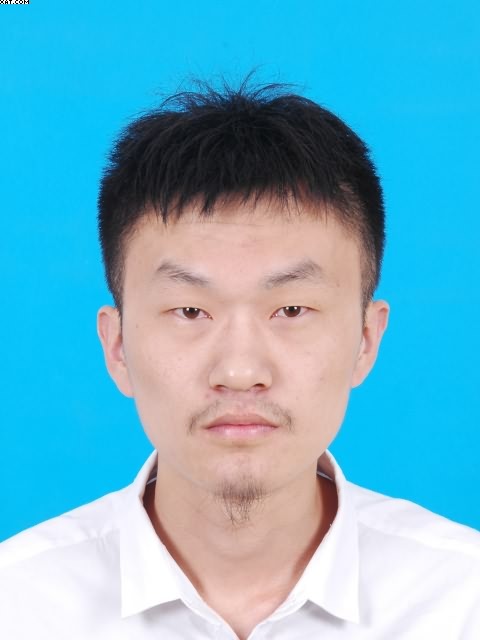}}]{Zhuang Liu}
received a Ph.D. degree from Beihang University in 2024. He works with the Engineering Research Center of Advanced Computer Application Technology, Ministry of Education, Beihang University. His main research interests are recommender systems and network representation learning.
\end{IEEEbiography}
\vspace{-10 mm}
\begin{IEEEbiography}[{\includegraphics[width=1in,height=1.25in,clip,keepaspectratio]{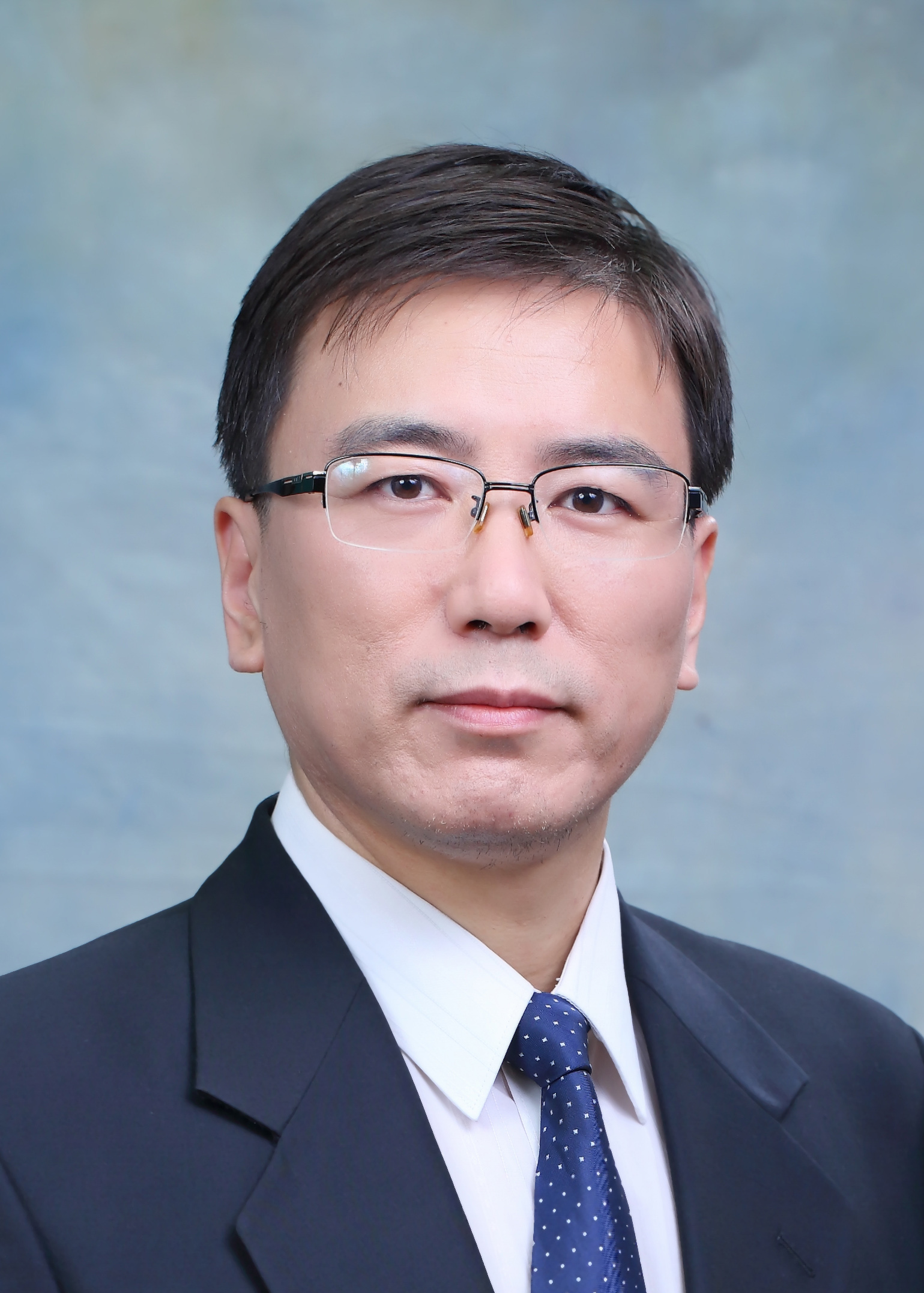}}]{Wenge Rong}
is a professor in School of Computer Science and Engineering, Beihang University, China. He received his PhD from University of Reading, UK, in 2010; MSc from Queen Mary College, University of London, UK, in 2003; and BSc from Nanjing University of Science and Technology, China, in 1996. He has many years of working experience as a senior software engineer in numerous research projects and commercial software products. His area of research covers machine learning, natural language processing, and information management.
\end{IEEEbiography}
\vspace{-10 mm}
\begin{IEEEbiography}[{\includegraphics[width=1in,height=1.25in,clip,keepaspectratio]{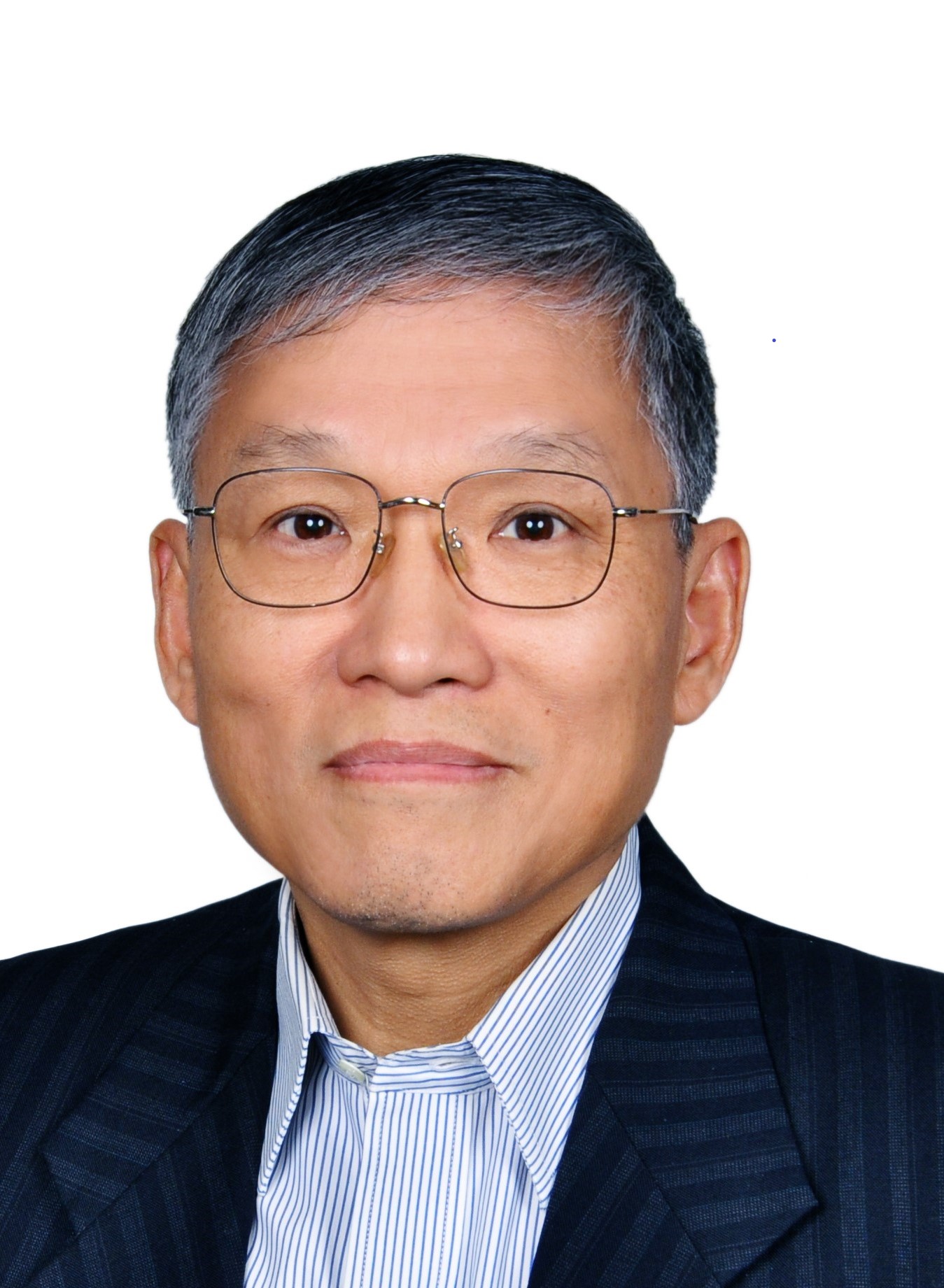}}]{Zhang Xiong}
received the B.S. degree from Harbin Engineering University in 1982 and the M.S. degree from Beihang University, China, in 1985. He is currently a Professor and a Ph.D. Supervisor with the School of Computer Science and Engineering, Beihang University, and the director of the Engineering Research Center of Advanced Computer Application Technology, Ministry of Education, Beihang University. His research interests and publications span from smart cities, digital ecuation, knowledge management, etc.
\end{IEEEbiography}

\vfill

\end{document}